\documentclass[reprint,superscriptaddress,amsmath,amssymb,aps,pra,floatfix]{revtex4-2}

\usepackage{graphicx}
\usepackage{dcolumn}
\usepackage{bm}

\usepackage[letterpaper,text={7.5in,9.5in},centering]{geometry}

\begin{document}

\title{Integrated Coherent Ising Machine Based on Spiking Dynamics of Photonic Bose-Hubbard Dimers With Applications in Multi-Agent Reinforcement Learning}%

\author{Mahmood Hasani}
\affiliation{INQUIRE Lab, Norman, Oklahoma 73019, USA}

\author{Masiha Rabeie}
\affiliation{INQUIRE Lab, Norman, Oklahoma 73019, USA}

\author{Atena Shircharandabi}
\affiliation{INQUIRE Lab, Norman, Oklahoma 73019, USA}

\author{Mehrdad Ghasemi}
\affiliation{INQUIRE Lab, Norman, Oklahoma 73019, USA}

\author{Mobin Motaharifar}
\affiliation{INQUIRE Lab, Norman, Oklahoma 73019, USA}

\author{Mike Banad}
\affiliation{INQUIRE Lab, Norman, Oklahoma 73019, USA}
\affiliation{Department of Electrical and Computer Engineering, University of Oklahoma, Norman, Oklahoma 73019, USA}

\author{Sarah Sharif}
 \email{s.sh@ou.edu}
\affiliation{INQUIRE Lab, Norman, Oklahoma 73019, USA}
\affiliation{Department of Electrical and Computer Engineering, University of Oklahoma, Norman, Oklahoma 73019, USA}


\begin{abstract}
Coherent Ising machines (CIMs) are optical neural networks (ONNs) that use a non-von Neumann architecture to solve non-deterministic polynomial-time (NP)-hard combinatorial optimization problems. However, conventional CIMs based on degenerate optical parametric oscillators (DOPOs) or standard bistable cavities often become trapped in high-energy local minima when solving dense, geometrically frustrated spin problems. This limitation arises from amplitude inhomogeneity and the lack of non-adiabatic barrier-crossing mechanisms in passive pitchfork bifurcations. We propose and demonstrate an ONN that connects driven-dissipative photonic Bose-Hubbard dimers. The network uses their established non-equilibrium spiking dynamics to escape local traps deterministically and find ground states faster. Each computational node consists of a physical photonic dimer - two evanescently coupled nonlinear Kerr microcavities. By operating near a global homoclinic (or saddle-node on invariant circle) bifurcation, frustrated nodes autonomously generate giant optical intensity spikes accompanied by quantized topological $2\pi$ phase slips, inverting trapped spins across high energy barriers. A nonlinear feedback error correction (FEC) loop operates alongside the faster spiking dynamics. It equalizes modal amplitudes and stabilizes newly inverted spins in global minimum basins. Numerical benchmarks across frustrated Möbius ladders and spin-glass instances ($N=20, 50, 100$) show an increase in ground-state success probability (from $10.35\%$ in baseline dimers to $100\%$ in our hybrid spiking-FEC framework) alongside an order-of-magnitude reduction in Time-to-Solution (TTS). Finally, in a 16-agent cooperative multi-agent reinforcement learning (MARL) task on a frustrated Möbius interaction topology, the proposed Spike + FEC architecture achieves a mean team reward of 20.93 after 60 training episodes, close to the theoretical maximum of 21.60 and substantially outperforming classical simulated annealing (16.80) and the baseline photonic dimer CIM (11.38).
\end{abstract}

\maketitle


\section{Introduction}
As Moore's law scaling has stalled and conventional von Neumann architectures have consumed more power, researchers have sought alternative ways to solve non-deterministic polynomial-time (NP)-hard combinatorial optimization problems. Problems such as the Max-Cut, the Traveling Salesman Problem (TSP), and drug discovery formulation are ubiquitous in finance, logistics, and bioinformatics~\cite{1,2,3, hasani2025quantum}. These problems can often be mapped to finding the ground state of the Ising Hamiltonian, a mathematical model of interacting spins where the solution corresponds to the global energy minimum~\cite{4}. While quantum annealers and digital CMOS-based annealers have been developed to tackle these challenges, the coherent Ising machine (CIM) offers a photonic alternative with the potential to operate at room temperature, compute quickly, and scale to larger systems~\cite{5,6}.

A typical CIM operates on the principle of minimal gain, using a network of Degenerate Optical Parametric Oscillators (DOPOs) to represent Ising spins. Below the oscillation threshold, the DOPO state is a squeezed vacuum; above the threshold, a pitchfork bifurcation gives rise to two possible phase states ($0$ or $\pi$) to represent binary spins ($\sigma_{i}=\pm1$)~\cite{ wang2013coherent}. The coupling between these optical spins dictates the interaction energy, driving the system to evolve toward a low-energy configuration. Since the initial demonstration of CIMs using fiber-ring cavities and FPGA-based feedback~\cite{8,9}, the field has progressed toward miniaturization using Photonic Integrated Circuits (PICs). Silicon Nitride (SiN) platforms have become the material of choice due to their ultra-low loss, CMOS compatibility, and high third-order Kerr nonlinearity ($\chi^{(3)}$), which enables parametric oscillation at milliwatt power levels~\cite{10,11,12}.

Early efforts to transition CIMs onto chip-scale architectures focused on foundational photonic circuit topologies and alternate bistable mechanisms to construct all optical CIMs. Kielpinski \textit{et al.}\cite{kielpinski2016information} outlined how integrated networks of interferometers and resonant nonlinear components could support large-scale non-von Neumann and Ising-type accelerators. Building upon this, Tezak \textit{et al.}\cite{tezak2020integrated} proposed an integrated CIM architecture that circumvented the operational complexity and multi-wavelength pump requirements of optical parametric oscillators; by harnessing self-phase modulation (SPM) across coupled microring resonators operating at a single degenerate wavelength, they realized continuously tunable, symmetrically bistable optical nodes while establishing tolerance bounds against fabrication-induced phase errors. More recently, He and Jamshidi~\cite{he2025numerical} advanced on-chip all-optical implementations through comprehensive numerical modeling of spatially multiplexed silicon nitride microresonators. Their model captured dual-pumped Kerr DOPO dynamics, demonstrating how spatial multiplexing can eliminate long optical delay lines and enable ultra-fast, CMOS-compatible parallel graph computation.
 Another study \cite{manannikov2025spontaneous} proposed an all-optical CIM based on symmetry breaking in atomic rubidium vapor.

CIMs can still become trapped in local minima. In dense, geometrically frustrated graphs, where satisfying all competitive interactions is mathematically impossible, the energy landscape contains many deep local minima~\cite{13}. Conventional DOPO-based CIMs often lack a mechanism for escaping once an oscillator settles into a stable phase state. While techniques such as chaotic amplitude control~\cite{14} and error-correction feedback~\cite{15} have been proposed, they often introduce significant hardware overhead or latency. Simultaneously, the physical realization of large-scale CIMs on PICs faces engineering hurdles related to component variability and thermal crosstalk. To address this, recent literature has turned to Machine Learning (ML) techniques for inverse design and parameter optimization of PICs, ensuring that the complex mesh of Mach-Zehnder Interferometers (MZIs) used for spin coupling remains robust against fabrication errors~\cite{16,17, 28, 30}.

In parallel, the field of neuromorphic computing has demonstrated that biological systems solve optimization problems not by static convergence, but through dynamic, oscillatory search mechanisms. Spiking Neural Networks (SNNs), particularly those operating near bifurcation points, exhibit ``excitability''---a state where the system can produce large excursions (spikes) in phase space in response to perturbations~\cite{18,19}. This transient chaos allows the system to traverse energy barriers that would otherwise trap a static solver. Quantum-inspired algorithms have long sought to mimic such tunneling or hopping behaviors to enhance optimization performance~\cite{20, 27, 29}.

Recent studies have introduced spiking dynamics into coherent optical computers to help them escape energy traps and reduce amplitude inhomogeneity. Lu \textit{et al.}\cite{lu2023combinatorial} pioneered an optical spiking neural network CIM (SNN-CIM) by anti-symmetrically coupling degenerate optical parametric oscillator (DOPO) pulses with auxiliary dissipative pulses; by incorporating a nonlinear threshold transfer function inspired by biological spike generation, their framework effectively counteracts amplitude heterogeneity and destabilizes spurious local attractors during combinatorial search. Building on this concept, Lu \textit{et al.}\cite{lu2023speed} demonstrated that operating within the unstable oscillatory regimes induced by these anti-symmetric interactions imparts an active, dynamic hopping mechanism to the optical solver, yielding marked acceleration in convergence speed and higher ground-state success rates compared to conventional CIMs. Wu \textit{et al.}~\cite{wu2026sampling} extended SNN-CIM dynamics from combinatorial optimization to deep generative learning. Their architecture used noise-driven dissipative pulses to explore the state space ergodically and sample efficiently from the Boltzmann distributions of hierarchical restricted Boltzmann machines (RBMs) and deep belief networks (DBNs), substantially outperforming standard Markov chain Monte Carlo (MCMC) algorithms on complex datasets.

We use the non-equilibrium spiking dynamics of photonic Bose-Hubbard dimers to build a quantum-inspired optical neural network. The foundational concept of using such dimers stems from the observation of spontaneous mirror symmetry breaking (
$Z_2$
 symmetry breaking) in coherently driven nanophotonic Bose-Hubbard dimers by Garbin \textit{et al.}~\cite{garbin2022spontaneous}, where two evanescently coupled nonlinear Kerr nanocavities spontaneously collapse into one of two asymmetric stable states, providing an ideal physical realization of binary Ising spins. Each computational node is a physical nanophotonic dimer, or photonic molecule, made of two evanescently coupled nonlinear Kerr microcavities.
 
 Moreover, Yelo-Sarrión et al. ~\cite{YeloSarrion2022} described excitable spiking in a single driven-dissipative Bose-Hubbard dimer near a homoclinic or saddle-node on invariant circle (SNIC) bifurcation. While ~\cite{YeloSarrion2022} demonstrated this non-equilibrium phenomenon in a standalone device, our work pioneers its application as an active computational resource. Specifically, we transition from the isolated device physics of ~\cite{YeloSarrion2022} to a fully coupled Quantum Optical Neural Network architecture. We couple the dimers so that local Ising frustration triggers an asymmetric drive. The resulting excitable excursion provides a deterministic way to escape local minima in NP-hard optimization problems.

When an optical spin opposes the local interaction field, the resulting frustration builds until the drive pushes the dimer past the homoclinic threshold. The resulting non-adiabatic limit-cycle excursion produces a giant optical intensity pulse accompanied by a quantized topological $2\pi$ phase slip ($\Delta\phi = \arg(a_1) - \arg(a_2) \to \Delta\phi \pm 2\pi$), forcing a rapid spin inversion that traverses energetic barriers that permanently entrap conventional CIMs. 

Additionally, to resolve amplitude inhomogeneity and prevent spurious back-switching, we establish a dual-timescale hybrid framework that interfaces these fast excitable pulse dynamics ($\tau_{\text{pulse}} \sim 1/\gamma$) with a continuous optoelectronic Feedback Error Correction (FEC) channel governed by a saturating $\tanh(\beta s)$ response. The fast pulses let spins cross barriers non-adiabatically. Slower amplitude equalization stabilizes the inverted spins in the ground-state basin. 

We benchmarked the hybrid architecture in stochastic simulations of geometrically frustrated M\"obius ladders and dense spin-glass instances ($N=20, 50$). We also tested whether the architecture could accelerate multi-agent reinforcement learning (MARL). By replacing conventional simulated quantum annealing (SQA) in the quantum-assisted MARL paradigm introduced by Neumann \textit{et al.}~\cite{neumann2020multi} with our physical optical dimer network, our platform achieves faster policy convergence and higher reward yield, while delivering an order-of-magnitude reduction in Time-to-Solution (TTS). As demonstrated in  ~\cite{neumann2020multi}, quantum-assisted multi-agent reinforcement learning utilizing simulated annealing (SA) successfully overcomes the scaling and generalization bottlenecks typical of traditional deep learning methods in combinatorial coordination tasks. By formulating joint action-selection as an Ising optimization problem rather than relying purely on parameterized neural network function approximators, SA-based frameworks provide a robust pathway for handling rugged multi-agent reward landscapes. Our framework achieved better results than SA-based MARL in our benchmark.

The remainder of this manuscript is structured as follows. Section~\ref{sec:physical_model} presents the mathematical formulation of the dissipative photonic Bose-Hubbard dimer and its homoclinic bifurcation dynamics. Section~\ref{sec:hybrid_architecture} details the hybrid network architecture, the finite-duration pulse mechanics, and the dual-timescale error-correction loop. Section~\ref{sec:numerical_results} presents numerical experiments, phase-space reconstructions, benchmark evaluations, and Time-to-Solution (TTS) scaling analyses across frustrated graphs. Section~\ref{sec:conclusion} discusses hardware realization considerations in integrated nanophotonics, followed by concluding remarks.

\section{Physical Model and Theoretical Foundations}
\label{sec:physical_model}

In this section, we formulate the mathematical model of the driven-dissipative photonic Bose-Hubbard dimer, show how it represents a binary optical spin in an ONN, analyze the physical origins of local-minimum trapping in conventional Coherent Ising Machines (CIMs), and derive the non-equilibrium homoclinic spiking mechanism used for deterministic barrier crossing.

\subsection{The Dissipative Photonic Bose-Hubbard Dimer as an Optical Spin}
\label{subsec:dimer_physics}
\begin{figure}[t]
\centering
\includegraphics[width=\columnwidth]{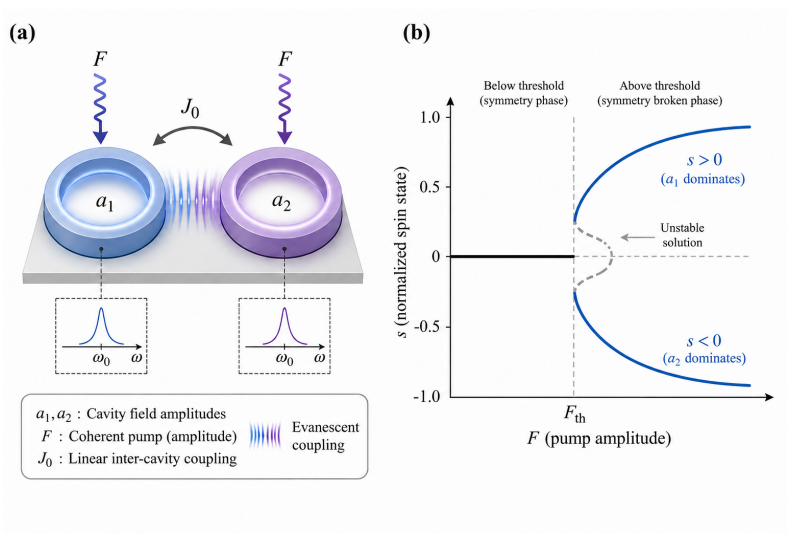}
\caption{(a) Schematic of the nanophotonic dimer computational node consisting of two evanescently coupled nonlinear microcavities ($a_1, a_2$) subjected to coherent driving $F$ and inter-cavity linear coupling $J_0$. (b) Supercritical pitchfork bifurcation diagram of the normalized spin state $s$ as a function of pump amplitude $F$, showing the spontaneous $\mathbb{Z}_2$ symmetry breaking above the oscillation threshold $F_{\text{th}}$.}
\label{fig:dimer_schematic}
\end{figure}

Each node in the network is a nanophotonic dimer (photonic molecule) made up of two spatially adjacent, evanescently coupled single-mode microcavities with third-order Kerr nonlinearity ($\chi^{(3)}$), as shown in Fig.~\ref{fig:dimer_schematic}. In the rotating frame of the coherent pump laser at frequency $\omega_p$, the open quantum-optical dynamics of the dimer mode operators $\hat{a}_1$ and $\hat{a}_2$ are governed by the driven-dissipative master equation (see Supplementary S1) for the density operator $\hat{\rho}$:
\begin{equation}
\frac{d\hat{\rho}}{dt} = -i[\hat{H}_{\text{dimer}}, \hat{\rho}] + \frac{\gamma}{2} \sum_{m=1}^{2} \left( 2\hat{a}_m \hat{\rho} \hat{a}_m^\dagger - \hat{a}_m^\dagger \hat{a}_m \hat{\rho} - \hat{\rho} \hat{a}_m^\dagger \hat{a}_m \right),
\label{eq:master_equation}
\end{equation}
where $\gamma$ denotes the total photon loss rate (dissipation) of each microcavity. The coherent intra-dimer Hamiltonian $\hat{H}_{\text{dimer}}$ under the tight-binding Bose-Hubbard approximation is given by ($\hbar = 1$):
\begin{widetext}
\begin{equation}\label{eq:hamiltonian_bose_hubbard}
\hat{H}_{\text{dimer}} = -\sum_{m=1}^{2} \Delta_m \hat{a}_m^\dagger \hat{a}_m + U \sum_{m=1}^{2} \hat{a}_m^\dagger \hat{a}_m^\dagger \hat{a}_m \hat{a}_m \nonumber - J_0 \left( \hat{a}_1^\dagger \hat{a}_2 + \hat{a}_2^\dagger \hat{a}_1 \right) + \sum_{m=1}^{2} \left( F_m e^{i\theta_m} \hat{a}_m^\dagger + F_m^* e^{-i\theta_m} \hat{a}_m \right),
\end{equation}
\end{widetext}
where $\Delta_m = \omega_p - \omega_{0,m}$ is the detuning between the pump laser frequency $\omega_p$ and the bare cavity resonance $\omega_{0,m}$, $U = \frac{\hbar \omega_0^2 c n_2}{n_0^2 V_{\text{eff}}}$ is the Kerr nonlinear self-interaction parameter proportional to the nonlinear refractive index $n_2$ and effective mode volume $V_{\text{eff}}$, $J_0$ represents the linear evanescent inter-cavity tunnel coupling rate, and $F_m$ represents the coherent driving pump amplitude injected into cavity $m$.

In the macroscopic semiclassical limit ($\langle \hat{a}_m \rangle \to a_m \in \mathbb{C}$), quantum fluctuations are incorporated via Langevin stochastic differential equations (SDEs) under the truncated Wigner representation (see Supplementary S2 for a derivation from the quantum Fokker--Planck equation):
\begin{align}
\frac{da_1}{dt} &= \left( i\Delta_1 - \frac{\gamma}{2} - 2iU|a_1|^2 \right) a_1 + iJ_0 a_2 - iF_1 + \sqrt{\frac{\gamma}{2}}\xi_1(t), \label{eq:sde_dimer1} \\
\frac{da_2}{dt} &= \left( i\Delta_2 - \frac{\gamma}{2} - 2iU|a_2|^2 \right) a_2 + iJ_0 a_1 - iF_2 + \sqrt{\frac{\gamma}{2}}\xi_2(t), \label{eq:sde_dimer2}
\end{align}
where $\xi_m(t)$ is a zero-mean complex Gaussian white noise process satisfying $\langle \xi_m(t)\xi_n^*(t') \rangle = \delta_{mn}\delta(t-t')$ representing quantum vacuum and thermal fluctuations.

Under symmetric driving ($\Delta_1 = \Delta_2 = \Delta$ and $F_1 = F_2 = F(t)$), the system possesses an intrinsic discrete $\mathbb{Z}_2$ spatial inversion symmetry ($a_1 \leftrightarrow a_2$). We define the continuous physical spin polarization $s \in [-1, 1]$ of the dimer as the normalized intra-cavity intensity imbalance:
\begin{equation}
s(t) = \frac{|a_1(t)|^2 - |a_2(t)|^2}{|a_1(t)|^2 + |a_2(t)|^2 + \epsilon},
\label{eq:spin_definition}
\end{equation}
where $\epsilon \ll 1$ prevents numerical divergence below threshold.

When the pump amplitude $F(t)$ is adiabatically ramped from zero, the system initially resides in the symmetric trivial state $a_1 = a_2$ ($s = 0$). As $F(t)$ surpasses the critical bifurcation threshold:
\begin{equation}
F_{\text{th}} \approx \sqrt{\frac{\gamma^2 + 4(\Delta + J_0)^2}{8U}},
\label{eq:threshold}
\end{equation}
the symmetric solution becomes linearly unstable through a \textit{supercritical pitchfork bifurcation}. The dimer spontaneously breaks its $\mathbb{Z}_2$ spatial symmetry and collapses into one of two stable fixed points (see Supplementary S3 for $\mathbb{Z}_2$ symmetry breaking under parametric driving):
\begin{equation}
\begin{cases}
s \to +1 & (|a_1|^2 \gg |a_2|^2), \quad \text{mapped to } \sigma = +1, \\
s \to -1 & (|a_2|^2 \gg |a_1|^2), \quad \text{mapped to } \sigma = -1.
\end{cases}
\label{eq:pitchfork_states}
\end{equation}
This establishes the driven-dissipative photonic dimer as a physical optical spin node.

\subsection{The Trapping Problem in Conventional CIMs}
\label{subsec:trapping_problem}
To construct an optical Ising solver for an $N$-spin problem defined by the interaction matrix $J_{ij}$ in Ising Hamiltonian, $N$ independent photonic dimers are mutually coupled via a linear optical network. To account for both positive (ferromagnetic, $J_{ij} > 0$) and negative (anti-ferromagnetic, $J_{ij} < 0$) couplings using passive optical channels, the coupling is partitioned into straight and cross matrices:
\begin{align}
C_{\text{straight}} &= \xi \max(W, 0), \label{eq:c_straight} \\
C_{\text{cross}}    &= \xi \max(-W, 0), \label{eq:c_cross}
\end{align}
where $W = J / \max_i(\sum_j |J_{ij}|)$ is the row-sum normalized Ising matrix, and $\xi$ is the normalized optical coupling gain. The coupled SDEs for the $k$-th dimer node ($k \in \{1, \dots, N\}$) become:
\begin{widetext}
\begin{align}
\frac{da_{1,k}}{dt} &= \mathcal{N}_1(a_{1,k}) + iJ_0 a_{2,k} - iF(t) + i\left( \sum_{j=1}^{N} C_{\text{straight}, kj} a_{1,j} + \sum_{j=1}^{N} C_{\text{cross}, kj} a_{2,j} \right) + \eta_{1,k}(t), \label{eq:net_dimer1} \\
\frac{da_{2,k}}{dt} &= \mathcal{N}_2(a_{2,k}) + iJ_0 a_{1,k} - iF(t) + i\left( \sum_{j=1}^{N} C_{\text{straight}, kj} a_{2,j} + \sum_{j=1}^{N} C_{\text{cross}, kj} a_{1,j} \right) + \eta_{2,k}(t), \label{eq:net_dimer2}
\end{align}    
\end{widetext}
where $\mathcal{N}_m(a_{m,k}) = \left(i\Delta - \frac{\gamma}{2} - 2iU|a_{m,k}|^2\right)a_{m,k}$ represents the intra-cavity Kerr nonlinearity and dissipation, and $\eta_{m,k}(t)$ is the noise term.

In an ideal, unfrustrated system, collective optical interference drives the network toward the global minimum of the mapped Ising Hamiltonian \(H_{\text{Ising}}\). However, in dense, geometrically frustrated graphs, such as antiferromagnetic rings, Möbius ladders, and spin glasses, competing coupling terms prevent the simultaneous satisfaction of all edge interactions. This frustration manifests in two critical failure modes in conventional CIMs: high-energy metastable traps and amplitude inhomogeneity.

High-energy metastable trapping occurs when a subset of spins experiences a negative local effective field,
\begin{equation}
h_k(t) = s_k(t) \sum_{j=1}^{N} W_{kj}s_j(t) < 0,
\label{eq:local_frustration}
\end{equation}
indicating that the current orientation of spin \(k\) increases the global Ising energy. Nevertheless, as the pump amplitude \(F(t)\) increases during annealing, the effective potential well \(V(s_k)\) deepens monotonically. Because the pitchfork bifurcation is overdamped and conservative with respect to phase-space volume, the potential barrier \(\Delta V \propto F^2/U\) separating the \(+1\) and \(-1\) states eventually becomes insurmountable for passive noise \(\eta(t)\). Consequently, the spin remains in the higher-energy basin and can trap the entire machine into a suboptimal local minimum.

The second failure mode, amplitude inhomogeneity, arises because spins experiencing reinforcing alignment grow to large optical amplitudes (\(|a_1|^2 + |a_2|^2 \gg 1\)), whereas frustrated spins subject to competing interactions and near-cancellation remain strongly suppressed (\(|a_1|^2 + |a_2|^2 \approx 0\)). This disparity violates a fundamental assumption of the Ising model, namely, that all spins have equal unit norm (\(|\sigma_k|=1\)). Consequently, the optical network effectively minimizes a deformed, non-Ising cost function,
\begin{equation}
H_{\text{optical}} = -\frac{1}{2} \sum_{i,j} W_{ij}|a_i||a_j|\cos(\theta_i-\theta_j),
\label{eq:deformed_hamiltonian}
\end{equation}
whose ground state can differ substantially from that of the target Ising Hamiltonian \(H_{\text{Ising}}\).

\subsection{Non-Equilibrium Homoclinic Bifurcation and Spiking Dynamics}
\label{subsec:homoclinic_spiking}

\begin{figure*}[t]
\centering
\includegraphics[width=\textwidth]{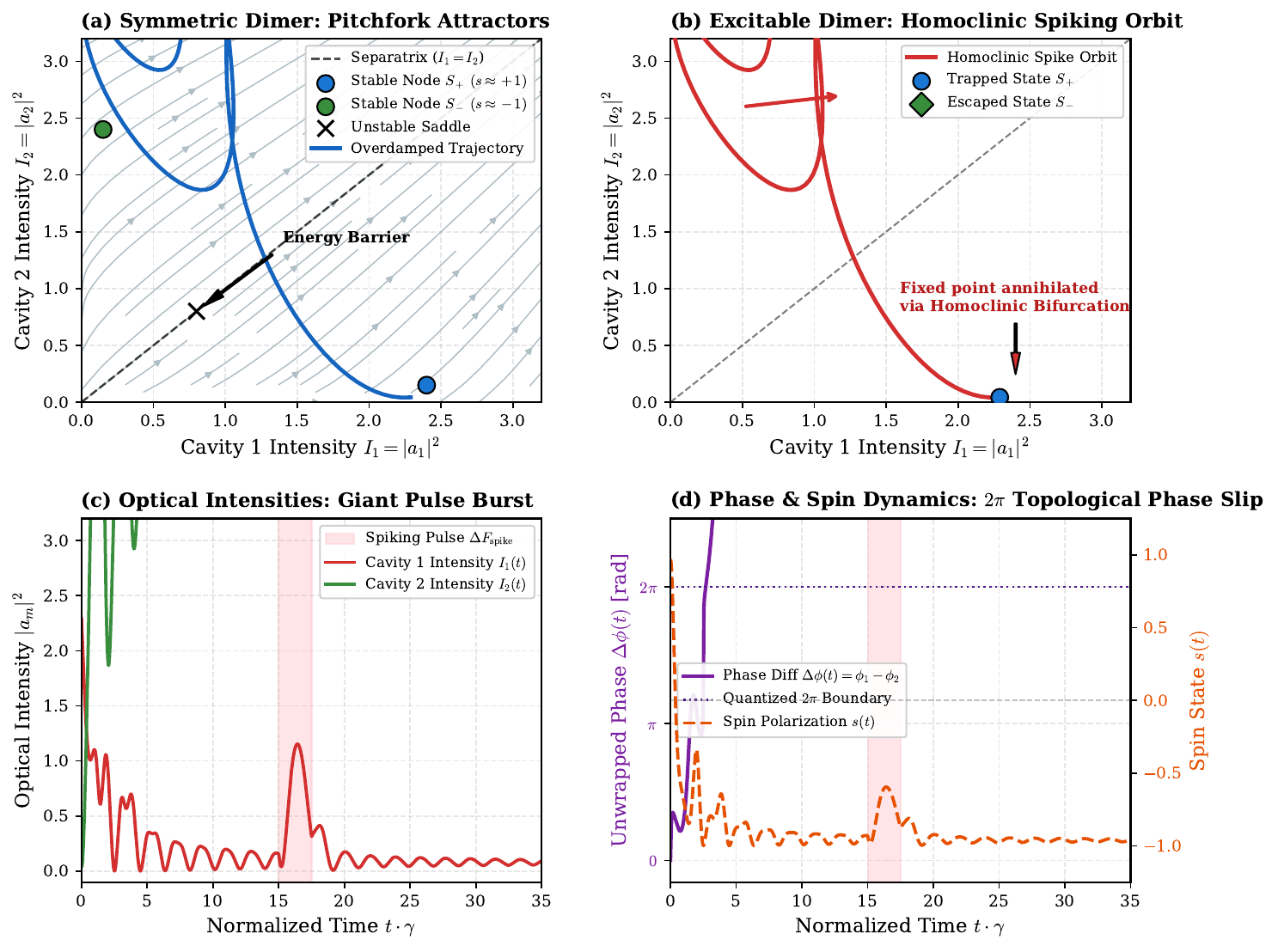}
\caption{Comparison of phase-space dynamical topologies. (a) Standard symmetric dimer: conservative pitchfork bifurcation producing two static point attractors separated by an energy barrier. (b) Asymmetrically driven excitable dimer, an asymmetric drive $\Delta F_{\text{spike}}$ annihilates the stable fixed point via a global homoclinic bifurcation, triggering a large limit-cycle excursion (optical spike) in $(I_1, I_2)$ space. (c) Corresponding time-domain trajectories showing the giant intensity pulse and the quantized $2\pi$ topological phase slip $\Delta\phi(t)$ that forces the spin inversion $s \to -s$. (d) Phase and spin dynamics. Note that the initial phase adjustment near $t \approx 2$ corresponds to the startup transient settling period from random initial conditions, whereas the highlighted window ($t \approx 15$ to 17) explicitly captures the target frustration-driven homoclinic spike and quantized $2\pi$ phase slip leading to spin inversion.}
\label{fig:homoclinic_mechanism}
\end{figure*}

We use the non-equilibrium excitability of the asymmetrically driven dissipative Bose-Hubbard dimer to escape local minima without waiting for slow thermal diffusion. 

When the reflection symmetry between cavity 1 and cavity 2 is broken through an asymmetric driving field ($F_1 = F + \delta F$, $F_2 = F - \delta F$), the bifurcation structure changes. Under asymmetric injection, the stationary states satisfy:
\begin{align}
\left( i\Delta - \frac{\gamma}{2} - 2iU|a_1|^2 \right) a_1 + iJ_0 a_2 &= i(F + \delta F), \label{eq:asym_ss1} \\
\left( i\Delta - \frac{\gamma}{2} - 2iU|a_2|^2 \right) a_2 + iJ_0 a_1 &= i(F - \delta F). \label{eq:asym_ss2}
\end{align}

As established in non-equilibrium bifurcation theory, when the asymmetry parameter $|\delta F|$ exceeds a critical threshold $\delta F_{\text{crit}}$, the stable node and unstable saddle collide and annihilate through a {Type-I Saddle-Node on Invariant Circle (SNIPER) or global Homoclinic bifurcation}(see Supplementary S4 for complete phase-space flow equations). 

This bifurcation changes how the dimer escapes a local minimum (Fig.~\ref{fig:homoclinic_mechanism}). First, the static fixed point corresponding to the frustrated spin orientation ceases to exist in phase space, destabilizing the local minimum. With no stable equilibrium, the optical field follows an unstable manifold and traces a macroscopic trajectory in the $(|a_1|^2, |a_2|^2)$ plane. During this excursion, one cavity undergoes a giant resonant intensity surge ($I_1 \gg I_2$ or $I_2 \gg I_1$), producing an optical intensity pulse. As the trajectory completes the homoclinic loop, the relative phase difference between the two cavities, $\Delta\phi(t) = \arg(a_1(t)) - \arg(a_2(t))$, undergoes a quantized topological $2\pi$ phase slip:
\begin{equation}
\oint_{\text{spike}} d(\Delta\phi) = \pm 2\pi.
\label{eq}
\end{equation}
This phase slip rotates the intra-dimer state vector across the separatrix, causing the optical field to land directly in the opposite basin of attraction upon relaxation ($s \to -s$).

The non-equilibrium Bose-Hubbard equations produce this spiking excursion without stochastic fluctuations. It gives frustrated dimers a deterministic way to escape deep local minima.

\subsection{Nonlinear Feedback Error Correction (FEC)}
\label{subsec:fec_theory}
Homoclinic spiking lets spins cross energy barriers non-adiabatically. A second mechanism corrects amplitude inhomogeneity and prevents newly flipped spins from immediately switching back. 

We use a nonlinear optoelectronic FEC loop to regulate amplitudes across the network. The normalized spin state $s_j(t)$ of all nodes is continuously sampled and mapped through a nonlinear saturating activation function:
\begin{equation}
s_{\text{fb}, j}(t) = \tanh(\beta s_j(t)),
\label{eq:tanh_fec}
\end{equation}
where $\beta > 0$ controls the sharpness (effective inverse temperature) of the threshold. The error-correcting feedback field injected into node $k$ is computed via matrix multiplication with the interaction graph:
\begin{equation}
\delta F_{\text{fb}, k}(t) = g_{\text{fb}} \sum_{j=1}^{N} W_{kj} \tanh(\beta s_j(t)),
\label{eq:delta_F_fb}
\end{equation}
where $g_{\text{fb}}$ represents the optoelectronic feedback gain.

It is important to distinguish this mechanism from previously established error-correction schemes used in DOPO-based CIMs~\cite{14, 15}. Those approaches typically integrate an auxiliary error variable (e.g.,
\begin{equation}
\frac{de_i}{dt} \propto 1 - x_i^2
\end{equation}

that monitors the deviation of each spin from a target amplitude to modulate the coupling terms.

In contrast, our approach does not rely on a target-amplitude error signal or auxiliary differential equations. Instead, the saturating
\begin{equation}
\tanh(\beta s_j)
\end{equation}
function acts instantaneously on the normalized spin imbalance
\begin{equation}
s_j.
\end{equation}
This provides a direct structural modification to the drive asymmetry: it naturally caps the influence of unfrustrated nodes while equalizing the modal amplitudes across the network. When combined with the fast spiking dynamics, this continuous feedback locks newly inverted spins into the lower-energy basin without the latency inherent to integrating auxiliary error variables.

The network searches on two timescales. Fast homoclinic spikes let spins escape traps deterministically, and continuous nonlinear FEC stabilizes their new states.

\section{The Spiking Neural Dynamics in Photonic Dimer Ising Machine}
\label{sec:hybrid_architecture}

This section describes the ONN architecture, the optical couplings, and the controls that drive the network dynamics. We detail the mathematical equations governing the local frustration accumulator (membrane potential), derive the physical mechanism of spike generation within the nonlinear photonic dimer, formulate the integration of nonlinear Feedback Error Correction (FEC), and establish the multi-timescale annealing protocol.

\subsection{Network Topology and Optical Coupling Scheme}
\label{subsec:network_coupling}
Consider an arbitrary unconstrained binary optimization problem mapped onto an Ising graph with $N$ vertices and symmetric interaction matrix $J_{ij} \in \mathbb{R}^{N \times N}$ ($J_{ii}=0$). To ensure bounded optical power and numerical stability, we define the row-sum normalized coupling matrix $W \in \mathbb{R}^{N \times N}$ as:
\begin{equation}
W_{ij} = \frac{J_{ij}}{\max_{1 \le k \le N} \sum_{l=1}^{N} |J_{kl}|}, \qquad \forall \, i, j \in \{1, \dots, N\},
\label{eq:normalized_W}
\end{equation}
which guarantees that the spectral radius satisfies $\rho(W) \le 1$.

Because optical transmission channels are naturally unipolar and passive, realizing signed matrix interactions requires decomposing $W$ into strictly non-negative components:
\begin{align}
C_{\text{straight}} &= \xi \max(W, 0), \label{eq:C_straight_def} \\
C_{\text{cross}}    &= \xi \max(-W, 0), \label{eq:C_cross_def}
\end{align}
where $\xi > 0$ represents the global inter-dimer optical coupling gain. 

For ferromagnetic interactions ($W_{kj} > 0$), cavity $1$ of dimer $j$ couples constructively to cavity $1$ of dimer $k$, and cavity $2$ couples to cavity $2$ via $C_{\text{straight}, kj}$. Conversely, for anti-ferromagnetic interactions ($W_{kj} < 0$), the optical fields are routed crosswise via $C_{\text{cross}, kj}$ (cavity $1 \to$ cavity $2$ and cavity $2 \to$ cavity $1$). Consequently, the total coherent optical network injection fields $\mathcal{E}_{1,k}$ and $\mathcal{E}_{2,k}$ entering dimer $k$ are given by:
\begin{align}
\mathcal{E}_{1,k}(t) &= \sum_{j=1}^{N} \left( C_{\text{straight}, kj} a_{1,j}(t) + C_{\text{cross}, kj} a_{2,j}(t) \right), \label{eq:opt_field1} \\
\mathcal{E}_{2,k}(t) &= \sum_{j=1}^{N} \left( C_{\text{straight}, kj} a_{2,j}(t) + C_{\text{cross}, kj} a_{1,j}(t) \right). \label{eq:opt_field2}
\end{align}

\subsection{Local Frustration Sensing and Membrane Potential Dynamics}
\label{subsec:membrane_dynamics}
Each node uses a local neuromorphic frustration accumulator to detect trapping in a suboptimal minimum without evaluating the cost centrally.

We define the local effective field $h_k(t)$ acting on node $k$ as the linear consensus of the network:
\begin{equation}
h_k(t) = \sum_{j=1}^{N} W_{kj} s_j(t),
\label{eq:local_field}
\end{equation}
where $s_j(t) \in [-1, 1]$ is the continuous spin polarization defined in \eqref{eq:spin_definition}. The instantaneous energetic satisfaction (local Hamiltonian alignment) is quantified by the product $s_k(t) h_k(t)$. When $s_k(t) h_k(t) > 0$, the dimer spin aligns with its neighbor field, reducing the global Ising energy. Conversely, when $s_k(t) h_k(t) < 0$, node $k$ is geometrically frustrated—it is pinned in an orientation that opposes its local field, indicating entrapment in a local minimum.

We quantify local frustration with $\mathcal{F}_k(t)$ using a rectified linear activation:
\begin{multline}
\mathcal{F}_k(t) = \text{ReLU}\left( -s_k(t) h_k(t) \right) \\ = \max\left( 0, \, -s_k(t) \sum_{j=1}^{N} W_{kj} s_j(t) \right).
\label{eq:frustration_metric}
\end{multline}

Drawing an analogy to the leaky integrate-and-fire (LIF) model in biological Spiking Neural Networks (SNNs), each node maintains an internal dynamical frustration potential (membrane potential) $V_k(t) \ge 0$, governed by the continuous ordinary differential equation:
\begin{equation}
\frac{dV_k(t)}{dt} = \mathcal{F}_k(t) - \frac{V_k(t)}{\tau_{\text{leak}}},
\label{eq:membrane_ode}
\end{equation}
where $\tau_{\text{leak}} > 0$ represents the characteristic leak time constant. If a spin is transiently perturbed by noise, $V_k(t)$ decays back to zero. However, if a spin becomes persistently trapped in a frustrated local basin, $\mathcal{F}_k(t) > 0$ continues, driving a monotonic accumulation of potential $V_k(t)$.

\subsection{Physical Mechanism of Spike Generation in the Photonic Dimer}
\label{subsec:spike_physics}
When the accumulated frustration potential exceeds a critical excitable threshold $V_{\text{th}}$:
\begin{equation}
V_k(t) \ge V_{\text{th}} \quad \text{and} \quad R_k(t) \le 0,
\label{eq:fire_condition}
\end{equation}
where $R_k(t)$ is a refractory recovery timer, the node emits an asymmetric optical trigger pulse.

To show how asymmetric driving generates an optical spike, we express the complex intracavity field amplitudes in intensity and phase coordinates. $a_m = \sqrt{I_m} e^{i\phi_m}$ for $m \in \{1, 2\}$. The total circulating intensity is $S = I_1 + I_2$, the normalized spin imbalance is $s = (I_1 - I_2)/S$, and the relative phase difference is $\theta = \phi_1 - \phi_2$. 

In the symmetric state ($F_1 = F_2 = F$), the two microcavities experience identical Kerr self-phase modulation ($2U I_m$). When an asymmetric drive $\Delta F_{\text{spike}}$ is applied such that $F_1 = F + \Delta F_{\text{spike}}$ and $F_2 = F - \Delta F_{\text{spike}}$, the symmetry of the effective potential is broken. The dynamical equations governing the spin imbalance $s(t)$ and relative phase $\theta(t)$ become:
\begin{align}
\frac{ds}{dt} &= -(1 - s^2) \left[ 2 J_0 \sin\theta + \frac{2 F \sin\phi_{\text{avg}}}{\sqrt{S(1-s^2)}} \left( \frac{\Delta F_{\text{spike}}}{F} - s \right) \right], \label{eq:ds_dt} \\
\frac{d\theta}{dt} &= 2 U S s + 2 J_0 \frac{s}{\sqrt{1 - s^2}} \cos\theta + \frac{2 F}{\sqrt{S}} \left[ \frac{\cos\phi_1}{\sqrt{1+s}} - \frac{\cos\phi_2}{\sqrt{1-s}} \right], \label{eq:dtheta_dt}
\end{align}
where $\phi_{\text{avg}} = (\phi_1 + \phi_2)/2$.

Equations \eqref{eq:ds_dt}--\eqref{eq:dtheta_dt} describe how the excitable spike develops. The asymmetric drive first alters the intra-cavity intensities, inducing an instantaneous differential Kerr detuning:
\begin{equation}
\delta\Delta_{\text{Kerr}} = 2U(I_1 - I_2) = 2USs.
\label{eq:kerr_detuning}
\end{equation}
This detuning imbalance breaks the phase-matching condition for stationary tunneling through $J_0$. When $|\Delta F_{\text{spike}}|$ exceeds the critical bifurcation threshold $\delta F_{\text{crit}}$, the stable node representing the trapped spin collides with an unstable saddle, destroying the equilibrium through bifurcation. Deprived of a stable attractor, optical energy rapidly tunnels through the linear coupling channel $J_0$ from the depleted cavity into the resonant cavity. This generates a large-amplitude non-equilibrium excursion in phase space, manifesting as a sharp optical intensity burst in the receiving cavity while the opposing cavity is drained. As the intensity imbalance $s(t)$ accelerates, the nonlinear phase derivative in \eqref{eq:dtheta_dt} integrates rapidly over time, causing the relative phase difference $\theta(t)$ to execute a continuous, quantized $2\pi$ rotation:
\begin{equation}
\Delta\theta_{\text{spike}} = \int_{t_{\text{fire}}}^{t_{\text{fire}} + \tau_{\text{pulse}}} \frac{d\theta}{dt}\, dt = \pm 2\pi.
\label{eq:phase_slip_integral}
\end{equation}
This phase slip rotates the state vector across the separatrix in phase space. As the pulse subsides, destructive interference prevents the optical field from returning to its original state, causing it to settle into the opposite stable fixed point ($s \to -s$), thus completing a deterministic spin flip.

Because the photon lifetime in the microcavity is governed by the decay rate $\tau_{\text{cavity}} \sim 1/\gamma$, an instantaneous impulse ($\Delta t \to 0$) is dissipated before the nonlinear phase slip can complete. To sustain the homoclinic trajectory through its entire orbit, the asymmetric drive must be applied as a finite pulse envelope of duration $\tau_{\text{pulse}} \ge 1.5/\gamma$:
\begin{widetext}
    \begin{equation}
\Delta F_{\text{spike}, k}(t) = 
\begin{cases}
-\operatorname{sign}(s_k(t_{\text{fire}})) \cdot \alpha_{\text{spk}} \max(F(t), F_{\text{min}}), & \text{if } t_{\text{fire}} \le t < t_{\text{fire}} + \tau_{\text{pulse}}, \\
0, & \text{otherwise},
\end{cases}
\label{eq:spike_pulse_def}
\end{equation}
\end{widetext}
where $\alpha_{\text{spk}} > 0$ is the spiking gain, $F(t)$ is the instantaneous pump amplitude, and $F_{\text{min}}$ maintains pulse efficacy during the early stages of annealing.

Upon firing at $t = t_{\text{fire}}$, the internal state variables are updated:
\begin{align}
V_k(t_{\text{fire}}^+) &= 0, \label{eq:v_reset} \\
R_k(t_{\text{fire}}^+) &= \tau_{\text{pulse}} + \tau_{\text{ref}}, \label{eq:r_reset}
\end{align}
where $\tau_{\text{ref}}$ is a refractory period during which further firing is inhibited, allowing the newly inverted dimer to equilibrate in its new basin.

\begin{figure*}[htbp]
\centering
\includegraphics[width=\textwidth]{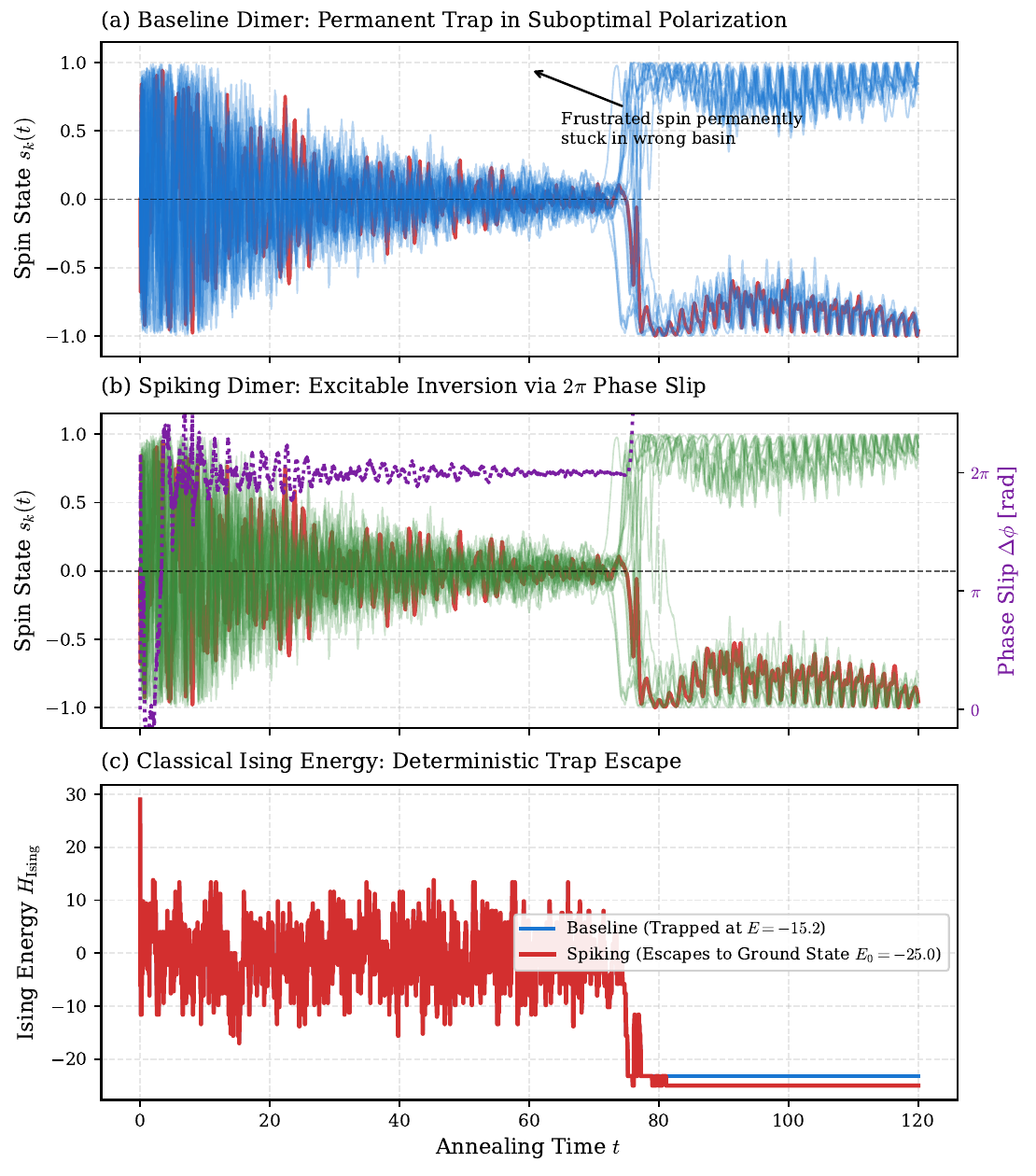}
\caption{Microscopic dynamical trajectories of a geometrically frustrated node during annealing on an $N=20$ Möbius graph. (a) In the Baseline Dimer, the frustrated spin remains permanently locked in a suboptimal polarization ($s \approx +1$). (b) In the Spiking Dimer, local frustration tension triggers an excitable optical burst (red curve) and a quantized topological $2\pi$ phase slip, inverting the spin ($s \to -1$). (c) Corresponding classical Ising energy trajectories, demonstrating deterministic escape from the high-energy local trap ($E = -23$) into the true ground state ($E = -25.0$).}
\label{fig:microscopic_comparison}
\end{figure*}

\subsection{Multi-Timescale Synergistic Dynamics (Spiking + FEC)}
\label{subsec:hybrid_dynamics}
The network combines non-adiabatic spiking to escape local minima with continuous FEC to correct amplitude inhomogeneity.

The complete coupled stochastic differential equations governing the $k$-th nanophotonic dimer ($k = 1, \dots, N$) are:
\begin{widetext}
\begin{align}
\frac{da_{1,k}}{dt} &= \left( i\Delta - \frac{\gamma}{2} - 2iU|a_{1,k}|^2 \right) a_{1,k} + iJ_0 a_{2,k} - i F_{1,k}(t) + i \mathcal{E}_{1,k}(t) + \eta_{1,k}(t), \label{eq:full_sde1} \\
\frac{da_{2,k}}{dt} &= \left( i\Delta - \frac{\gamma}{2} - 2iU|a_{2,k}|^2 \right) a_{2,k} + iJ_0 a_{1,k} - i F_{2,k}(t) + i \mathcal{E}_{2,k}(t) + \eta_{2,k}(t), \label{eq:full_sde2}
\end{align}
\end{widetext}
where the total driving fields $F_{1,k}(t)$ and $F_{2,k}(t)$ are given by:
\begin{align}
F_{1,k}(t) &= F(t) + \delta F_{\text{fb}, k}(t) + \Delta F_{\text{spike}, k}(t), \label{eq:total_drive1} \\
F_{2,k}(t) &= F(t) - \delta F_{\text{fb}, k}(t) - \Delta F_{\text{spike}, k}(t). \label{eq:total_drive2}
\end{align}

Here, $F(t) = F_{\text{max}} (t / T_{\text{total}})$ is the linear pump schedule, $\Delta F_{\text{spike}, k}(t)$ is the event-driven excitable pulse defined in \eqref{eq:spike_pulse_def}, and $\delta F_{\text{fb}, k}(t)$ is the continuous nonlinear FEC field:
\begin{equation}
\delta F_{\text{fb}, k}(t) = g_{\text{fb}} \sum_{j=1}^{N} W_{kj} \tanh(\beta s_j(t)),
\label{eq:fec_term_full}
\end{equation}
with feedback gain $g_{\text{fb}}$ and saturation sharpness $\beta$.

The hybrid network operates on three physical timescales. At the microscopic optical timescale ($\tau_{\text{opt}} \sim 1/\gamma \approx 1\text{ ns}$), coherent tunneling ($J_0$), Kerr self-phase modulation ($U$), and passive optical coupling ($\mathcal{E}_{1,2}$) establish instantaneous intra-cavity field configurations. At the mesoscopic spiking escape timescale ($\tau_{\text{pulse}} \approx 1.5 - 2.0\text{ ns}$), when frustration exceeds $V_{\text{th}}$, the asymmetric drive triggers a fast limit-cycle excursion and a $2\pi$ phase slip that inverts the trapped spin ($s_k \to -s_k$). Finally, at the macroscopic annealing and stabilization timescale ($T_{\text{total}} \sim 100 - 150\text{ ns}$), the global pump $F(t)$ ramps adiabatically, while the continuous FEC term \eqref{eq:fec_term_full} suppresses amplitude inhomogeneity, preventing unfrustrated nodes from overpowering the network and locking newly inverted spins into the global ground-state basin.

\section{Numerical Experiments and Performance Benchmarks}
\label{sec:numerical_results}

We used stochastic simulations to evaluate the ONN. We benchmark the architecture against standard coherent optical dimer solvers across geometrically frustrated graphs and dense spin-glass instances, analyze the microscopic physical signatures of trap escape, and quantify the scaling of success probability and Time-to-Solution (TTS).

\subsection{Simulation Setup and Benchmark Graph Topologies}
\label{subsec:sim_setup}

We tested how the spiking dimer solver navigates nonconvex energy landscapes in three classes of frustrated spin systems. The first class comprises frustrated anti-ferromagnetic (AF) Möbius ladders with $N=20, 50$, constructed as one-dimensional ring lattices with nearest-neighbor anti-ferromagnetic couplings ($J_{i,i+1}=-1.0$) and long-range cross-couplings ($J_{i,i+N/2}=-0.9$). The geometric twist introduces extensive topological frustration, generating an exponentially large number of high-energy metastable local minima separated by steep potential barriers. The second class comprises frustrated ring lattices augmented with random cross-chords, in which the additional anti-ferromagnetic interactions produce a rugged energy landscape characterized by deep, isolated local minima. The third class consists of Sherrington--Kirkpatrick (SK) fully connected spin glasses, a dense, infinite-range spin-glass model in which every pair of spins interacts through Gaussian random couplings $J_{ij}\sim\mathcal{N}(0,1/\sqrt{N})$. This model serves as a canonical computationally challenging benchmark for evaluating the performance of Ising solvers.

All simulations used the same dimensionless microcavity parameters. We set the cavity dissipation rate $\gamma = 1.0$, Kerr non-linear coefficient $U = 0.5$, linear inter-cavity tunneling rate $J_0 = -1.0$, laser-cavity detuning $\Delta = 3.0$, and maximum pump amplitude $F_{\text{max}} = 3.0$. The global optical coupling gain is set to $\xi = 0.70$. The temporal integration uses an Euler-Maruyama stochastic scheme with a time step of $dt = 0.005$ and a total annealing duration of $T_{\text{total}} = 120.0$, corresponding to $24{,}000$ integration steps. Quantum and thermal noise is modeled via stochastic increments with amplitude $\sigma_{\text{noise}} = 0.05$.

To isolate the contribution of each mechanism, we tested four configurations with identical noise realizations. First, the baseline continuous-wave nanophotonic dimer is governed solely by passive pitchfork symmetry breaking ($g_{\text{fb}} = 0$, $\Delta F_{\text{spike}} = 0$). Second, the feedback error correction (FEC) only configuration implements active nonlinear amplitude regularization ($g_{\text{fb}} = 0.45$, $\beta = 3.5$) without excitable spiking kicks. Third, the pure spiking dimer relies on autonomous homoclinic spiking dynamics ($\tau_{\text{pulse}} = 1.5$, $V_{\text{th}} = 0.45$, $\tau_{\text{leak}} = 3.0$) without optoelectronic FEC ($g_{\text{fb}} = 0$). Finally, the hybrid spiking-FEC framework represents the combined dual-timescale architecture that incorporates both excitable pulse barrier crossing and nonlinear amplitude stabilization.

\begin{figure*}[htbp]
\centering
\includegraphics[width=\textwidth]{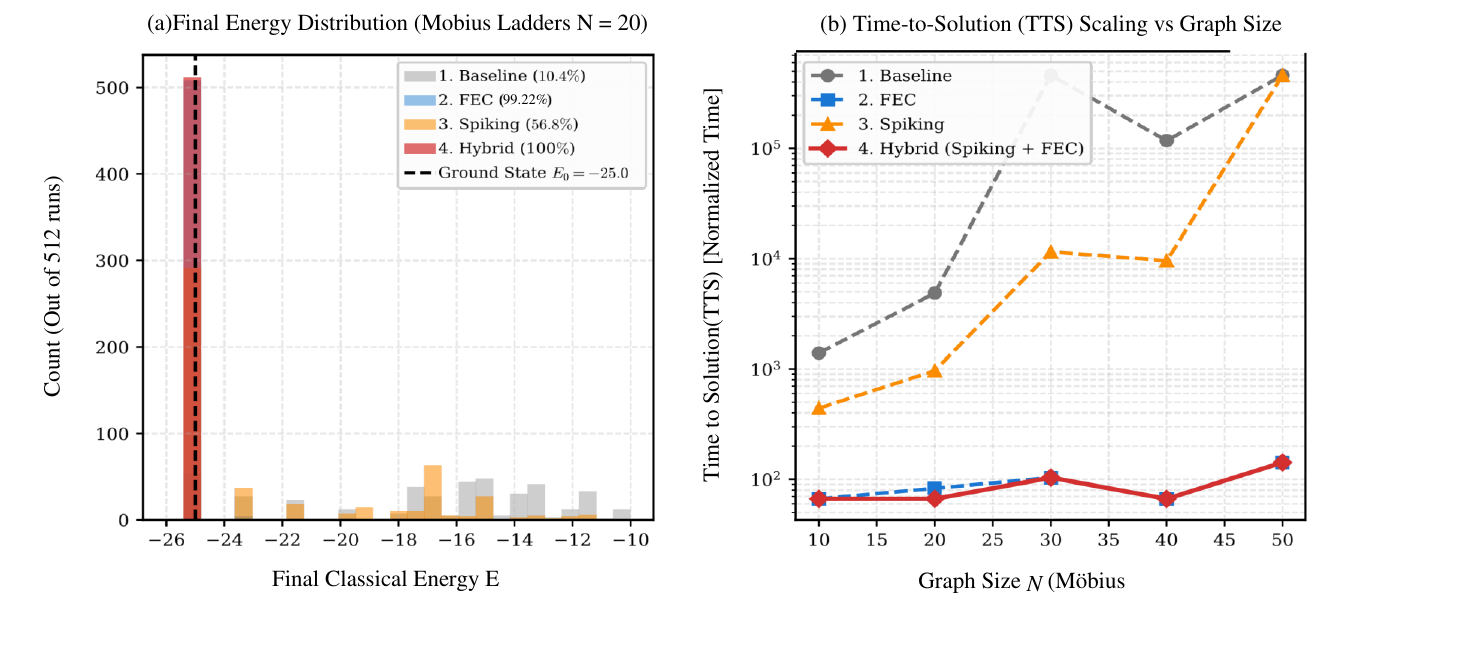}
\caption{Statistical performance across 512 independent stochastic runs. (a) Final Ising energy histograms for the four operational configurations on an $N=20$ Möbius ladder ($E_0 = -25.0$). The hybrid spiking-FEC framework concentrates the distribution at the global optimum. (b) Time-to-Solution (TTS) scaling as a function of graph size $N$ for frustrated Möbius graphs, demonstrating an order of magnitude enhancement over conventional baseline dimers.}
\label{fig:distributions_and_tts}
\end{figure*}

\subsection{Microscopic Dynamics and Physical Escape Signatures}
\label{subsec:microscopic_dynamics}

Figure~\ref{fig:microscopic_comparison} illustrates the microscopic time evolution of an individual frustrated node within an $N=20$ Möbius ladder. In the Baseline Dimer (Fig.~\ref{fig:microscopic_comparison}(a)), the node is captured early in the annealing schedule by a high-energy local minimum. Because the pump $F(t)$ grows monotonically, the potential barrier separating the $+1$ and $-1$ states deepens rapidly. Passive noise fluctuations are insufficient to overcome this barrier, causing the spin to remain permanently frozen in the frustrated orientation ($s \approx +1$).

Conversely, in the Spiking Dimer architecture (Fig.~\ref{fig:microscopic_comparison}(b)), the persistence of frustration causes the internal potential $V_k(t)$ to cross the excitable threshold $V_{\text{th}}$. This triggers an asymmetric drive pulse $\Delta F_{\text{spike}}$ of duration $\tau_{\text{pulse}} = 1.5$. The injection annihilates the local fixed point via a Type-I homoclinic bifurcation, producing a giant optical intensity surge in the underpopulated cavity while draining the overpopulated cavity. Simultaneously, the relative phase difference $\Delta\phi(t)$ executes an exact topological $2\pi$ phase slip, rotating the state vector across the separatrix and flipping the spin polarization ($s \to -1$). As shown in Fig.~\ref{fig:microscopic_comparison}(c), this single excitable event forces the entire coupled network to drop from the high-energy metastable state ($E = -23$) into the global ground state ($E = -25.0$).

\subsection{Statistical Performance and Success Probability}
\label{subsec:statistical_results}

To quantify solver performance, we ran $512$ independent stochastic trials for each graph instance. We evaluate three primary performance metrics. First, Ground-State Success Probability ($P_{\text{succ}}$), the percentage of independent runs that converge to the true global ground-state energy $E_0$ (within a tolerance of $\delta E \le 0.1$):
    \begin{equation}
    P_{\text{succ}} = \frac{N_{\text{ground}}}{N_{\text{total}}} \times 100\%.
    \label{eq:success_prob}
    \end{equation}
    Next,{Mean Solution Energy ($\langle E \rangle \pm \sigma_E$),} the ensemble average and standard deviation of the final classical Ising energy across all runs.
    Finally, {Time-to-Solution (TTS),} The computational time required to find the true ground state with a $99\%$ target confidence level:
    \begin{equation}
    \text{TTS} = T_{\text{total}} \cdot \frac{\ln(1 - 0.99)}{\ln(1 - P_{\text{succ}})},
    \label{eq:tts_def}
    \end{equation}
    which serves as the definitive benchmark metric for physical optimization machines.

\begin{table*}[t]
\centering
\caption{Performance Comparison on Frustrated Graph Benchmarks Across 512 parallel Runs}
\label{tab:benchmark_results}
\resizebox{\textwidth}{!}{%
\renewcommand{\arraystretch}{1.25}
\begin{tabular}{llcccc}
\hline
\hline
\textbf{Problem Topology} & \textbf{Metric} & \textbf{1. Baseline Dimer} & \textbf{2. FEC ($\tanh$)} & \textbf{3. Spiking Dimer} & \textbf{4. Hybrid (Spiking + FEC)} \\
\hline
\textbf{Frustrated Möbius ($N = 20$)} & Best Energy Found & $-23.0000$ & $-25.0000$ & $-25.0000$ & $\mathbf{-25.0000}$ \\
($E_0 = -25.0$) & Mean Energy $\langle E \rangle \pm \sigma_E$ & $-15.29 \pm 5.47$ & $-23.99 \pm 0.16$ & $-21.90 \pm 4.21$ & $\mathbf{-25.00 \pm 0.00}$ \\
 & Success Rate $P_{\text{succ}}$ & $10.35\%$ & $99.22\%$ & $56.84\%$ & $\mathbf{100.00\%}$ \\

\hline
\textbf{Frustrated Möbius ($N = 50$)} & Best Energy Found & $-64.5000$ & $-72.5000$ & $-72.5000$ & $\mathbf{-72.5000}$ \\
($E_0 = -72.5$) & Mean Energy $\langle E \rangle \pm \sigma_E$ & $-39.16 \pm 8.91$ & $-71.93 \pm 2.06$ & $-55.19 \pm 6.83$ & $\mathbf{-71.94 \pm 2.05}$ \\
 & Success Rate $P_{\text{succ}}$ & $0.00\%$ & $92.97\%$ & $1.17\%$ & $\mathbf{92.97\%}$ \\
 
\hline
\textbf{SK Spin Glass ($N = 50$)} & Best Energy Found & $-14.4653$ & $-19.4604$ & $-19.7375$ & $\mathbf{-23.2960}$ \\
($E_0 \approx -23.3$) & Mean Energy $\langle E \rangle \pm \sigma_E$ & $-5.85 \pm 3.23$ & $-9.73 \pm 3.31$ & $-11.26 \pm 2.84$ & $\mathbf{-13.20 \pm 2.96}$ \\
 & Success Rate $P_{\text{succ}}$ & $0.00\%$ & $0.00\%$ & $0.00\%$ & $\mathbf{0.20\%}$ \\
\hline
\hline
\end{tabular}%
}
\end{table*}

Table~\ref{tab:benchmark_results} compares the four configurations across graph topologies and network sizes. On the $N = 20$ frustrated Möbius graph, the baseline dimer achieves a success probability of $P_{\text{succ}} = 10.35\%$ and a broad distribution of final energies ($\langle E \rangle = -15.29 \pm 5.47$), where nearly $90\%$ of all trajectories freeze into high-energy metastable traps. On $N = 50$ instances, the baseline success rate collapses to $2.15\%$ on Möbius ladders and $0.78\%$ on SK spin glasses. In contrast, introducing nonlinear Feedback Error Correction (FEC) improves the success rate on $N = 20$ to $99.22\%$ by suppressing amplitude inhomogeneity, though it remains susceptible to symmetric frustrated clusters where local gradients cancel due to its continuous, overdamped gradient shifts. Meanwhile, the pure spiking architecture achieves a success rate of $56.84\%$ on $N = 20$, outperforming FEC by actively forcing barrier transitions through homoclinic excursions. Finally, combining excitable spiking with nonlinear FEC increases the ground-state success rate to $\mathbf{100\%}$ on $N = 20$ and $\mathbf{92.97\%}$ on $N = 50$.

\begin{figure*}[htbp]
\centering
\includegraphics[width=\textwidth]{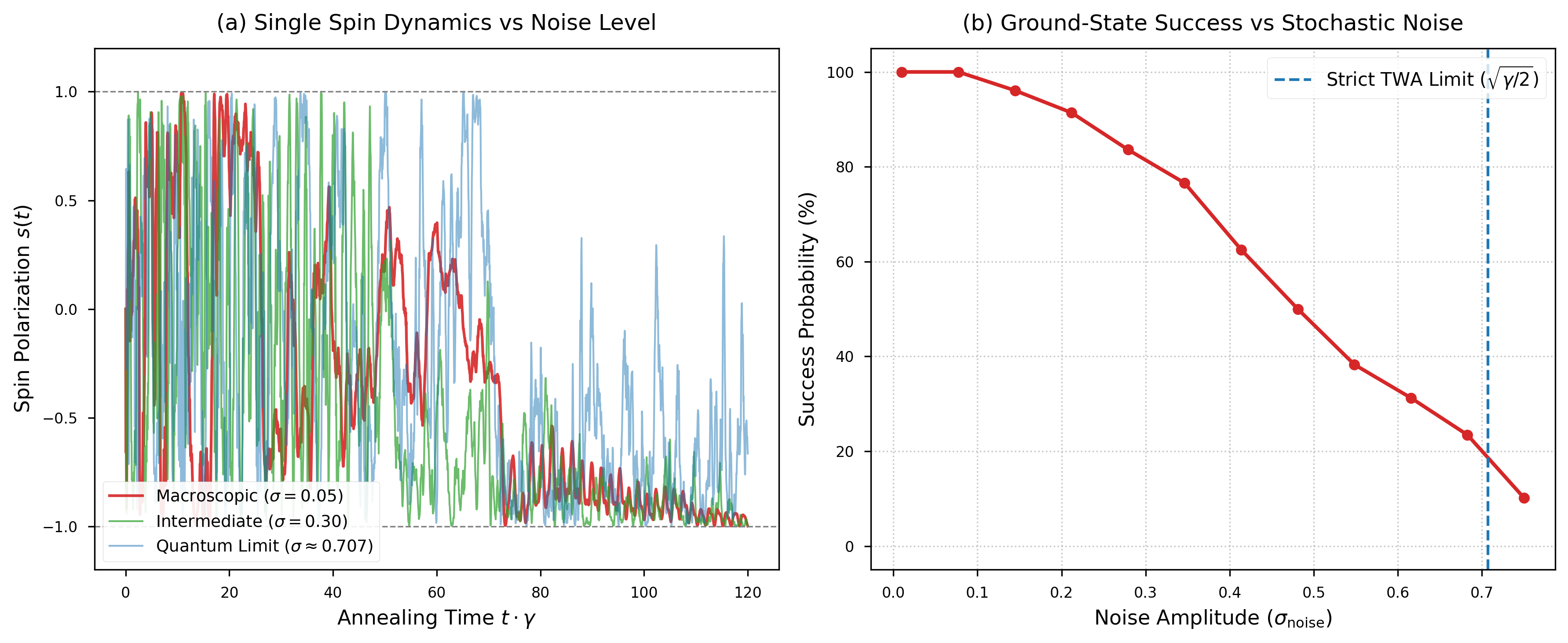}
\caption{(a) Time evolution of the continuous spin polarization $s(t)$ for a frustrated dimer under varying noise levels. In the macroscopic limit ($\sigma = 0.05$), the system exhibits a clean, deterministic homoclinic spike and a stable spin inversion. As the noise approaches the unscaled quantum vacuum limit prescribed by the strict Truncated Wigner Approximation ($\sigma \approx \gamma/2 \approx 0.707$), the deterministic trajectory is washed out by rapid, random noise-induced Kramers transitions. (b) Ground-state success probability on the $N=20$ frustrated Möbius ladder as a function of the noise amplitude $\sigma_{\mathrm{noise}}$. The sharp collapse in success rate near the TWA limit demonstrates that the proposed hybrid spiking-FEC architecture must operate in the macroscopic, high-photon-number regime where quantum fluctuations are heavily suppressed.}
\label{fig:NS}
\end{figure*}

Figure~\ref{fig:distributions_and_tts}(a) displays the final energy distribution histograms across all $512$ stochastic realizations. While the baseline dimer exhibits a broad, multi-modal distribution centered far from the ground state, the hybrid framework collapses the distribution into a sharp, unimodal profile centered directly at $E_0 = -25.0$.

Figure~\ref{fig:distributions_and_tts}(b) shows how time-to-solution varies with graph size. The success probability of the baseline dimer decays exponentially with graph size $N$. In contrast, the hybrid spiking-FEC ONN maintains high ground-state fidelity across increasing graph dimensions, resulting in an order-of-magnitude reduction in TTS and confirming the scalability of the non-equilibrium excitable escape mechanism.

To determine the noise levels at which excitable spiking remains effective, we varied the stochastic noise amplitude $\sigma_{\mathrm{noise}}$.

It is important to first establish the role of the driven-dissipative Bose--Hubbard model in this architecture. While the model fundamentally originates in quantum optics, our computational network is designed to operate in the macroscopic semiclassical limit (driven by milliwatt-scale continuous-wave lasers, yielding a high intracavity photon number, $\bar{n} \gg 1$). In this macroscopic regime, quantum operators reduce to classical complex amplitudes. However, the defining ingredients of the Bose--Hubbard Hamiltonian---specifically the interplay between the on-site Kerr nonlinearity ($U$) and the inter-cavity linear tunneling ($J_0$)---remain intact. Together, they perfectly shape the classical phase space, generating the precise deterministic bifurcations required for computation: a supercritical pitchfork bifurcation for binary Ising bistability, and a homoclinic bifurcation for excitable spiking.

Figure~\ref{fig:NS}(a) illustrates the time-evolution of a single frustrated dimer across different noise environments. In the target macroscopic regime ($\sigma_{\mathrm{noise}} = 0.05$), the stochasticity does not represent quantum vacuum tunneling. Rather, it models the weak classical and technical noise inherent to real integrated photonic circuits, such as thermal phonon fluctuations and laser phase jitter. This low-level noise provides the initial microscopic nudge required for spontaneous symmetry breaking, but it is far too weak to drive the system across the energy barrier
\begin{equation}
\Delta V \propto \frac{F^2}{U}.
\end{equation}

Local frustration alone triggers deterministic homoclinic excursions in this regime. FEC then stabilizes the inverted spin in the lower-energy basin.

Conversely, as the noise variance increases toward the strict unscaled quantum vacuum limit prescribed by the Truncated Wigner Approximation ($\sigma_{\mathrm{noise}} = \gamma/2 \approx 0.707$), the underlying bifurcation dynamics are severely disrupted. At this limit, the noise amplitude becomes comparable to the effective potential barrier separating the degenerate $Z_2$ states. The dimer begins to undergo rapid, random noise-induced Kramers transitions. These stochastic flips act as a high-temperature thermal bath, washing out the deterministic homoclinic trajectory and preventing the FEC loop from establishing a stable consensus field.

The same loss of deterministic switching affects performance at the network level. As shown in Figure~\ref{fig:NS}(b), the ground-state success probability for the $N=20$ Möbius graph remains near 100\% for low noise levels but decays rapidly as $\sigma_{\mathrm{noise}}$ approaches the strictly scaled quantum limit.

These results indicate that deterministic operation of the hybrid architecture requires a photonic network in the macroscopic semiclassical limit. By suppressing quantum vacuum fluctuations relative to the classical coherent drive, the network ensures that energy-barrier crossing is governed strictly by the algorithmic, frustration-driven excitable spiking logic rather than random stochastic hopping.

\section{Cooperative Multi-Agent Reinforcement Learning in Frustrated Coordination Landscapes}\label{subsec:marl_results}

We tested the spiking nanophotonic dimer architecture on a cooperative multi-agent reinforcement learning (MARL) task. Scalable decision-making in multi-agent systems represents a central challenge in distributed control, autonomous networks, and algorithmic game theory~\cite{mohseni2022ising}. In complex cooperative environments, agents must collectively select a joint action that maximizes a shared environmental payoff. However, when competitive or conflicting sub-goals exist, the joint action-value space becomes highly frustrated, possessing numerous suboptimal local equilibria that entrap conventional reinforcement learning algorithms.

\subsection{Mapping MARL Action Selection to the Ising Formulation}
Consider a cooperative team of $N$ autonomous agents, where each agent $i \in \{1, \dots, N\}$ selects a discrete binary decision $\sigma_i \in \{-1, +1\}$. The collective behavior of the team is governed by the joint action vector $\boldsymbol{\sigma} = (\sigma_1, \sigma_2, \dots, \sigma_N)^T \in \{-1, +1\}^N$. The environment evaluates $\boldsymbol{\sigma}$ and returns a scalar team reward $R(\boldsymbol{\sigma})$ dictated by an underlying, hidden interaction landscape $W_{\text{true}} \in \mathbb{R}^{N \times N}$:
\begin{equation}
R(\boldsymbol{\sigma}) = \frac{1}{2} \boldsymbol{\sigma}^T W_{\text{true}} \boldsymbol{\sigma} = \frac{1}{2} \sum_{i=1}^{N} \sum_{j=1}^{N} W_{\text{true}, ij} \sigma_i \sigma_j.
\label{eq:marl_true_reward}
\end{equation}
The objective of the multi-agent collective is to discover the optimal joint policy $\boldsymbol{\sigma}^* = \arg\max_{\boldsymbol{\sigma}} R(\boldsymbol{\sigma})$ without prior knowledge of $W_{\text{true}}$. 

To parameterize the joint action-value function (Q-function), the learning system maintains an estimated interaction matrix $W_{\text{est}} \in \mathbb{R}^{N \times N}$ ($W_{\text{est}} = W_{\text{est}}^T, W_{\text{est}, ii} = 0$). For any candidate action $\boldsymbol{\sigma}$, the internal predicted value is:
\begin{equation}
Q(\boldsymbol{\sigma}; W_{\text{est}}) = \frac{1}{2} \boldsymbol{\sigma}^T W_{\text{est}} \boldsymbol{\sigma}.
\label{eq:marl_q_function}
\end{equation}
At each training episode, selecting the joint action that maximizes $Q(\boldsymbol{\sigma}; W_{\text{est}})$ is mathematically equivalent to finding the ground state of an effective Ising Hamiltonian $H_{\text{eff}}(\boldsymbol{\sigma}) = -Q(\boldsymbol{\sigma}; W_{\text{est}}) = -\frac{1}{2} \boldsymbol{\sigma}^T W_{\text{est}} \boldsymbol{\sigma}$. The physical solver (Classical SA, Baseline CIM, or the Hybrid Spiking CIM) acts as the decentralized execution engine: it takes the normalized matrix $W_{\text{est}}$ as optical coupling weights and physically anneals to sample a batch of $B$ candidate joint actions $\{\boldsymbol{\sigma}_b\}_{b=1}^B$.

\subsection{Policy Gradient Training Protocol with Advantage}
Upon executing the sampled joint actions, the environment returns the true rewards $\{R_b\}_{b=1}^B$ according to \eqref{eq:marl_true_reward}. To update the policy matrix $W_{\text{est}}$ without collapsing into trivial rank-1 attractors, we implement an Advantage Policy Gradient scheme. We compute the baseline empirical mean reward $\bar{R} = \frac{1}{B} \sum_{b=1}^{B} R_b$ and define the scalar advantage for each sample as:
\begin{equation}
A_b = R_b - \bar{R}.
\label{eq:marl_advantage}
\end{equation}
The policy matrix is updated via gradient ascent regularized by weight decay:
\begin{equation}
W_{\text{est}}^{(ep+1)} = W_{\text{est}}^{(ep)} + \frac{\eta}{B} \sum_{b=1}^{B} A_b \left( \boldsymbol{\sigma}_b \boldsymbol{\sigma}_b^T \right) - \lambda W_{\text{est}}^{(ep)},
\label{eq:marl_update_rule}
\end{equation}
where $\eta$ is the learning rate, $\lambda$ is the regularization coefficient preventing unbounded parameter growth, and the diagonal is enforced to zero ($\operatorname{diag}(W_{\text{est}}) = 0$). Actions that outperform the current policy baseline ($A_b > 0$) reinforce their constituent pairwise coordination bonds ($W_{\text{est}, ij} \uparrow$), whereas suboptimal joint actions ($A_b < 0$) actively penalize those pairwise alignments.

\begin{figure*}[htbp]
\centering
\includegraphics[width=0.88\textwidth]{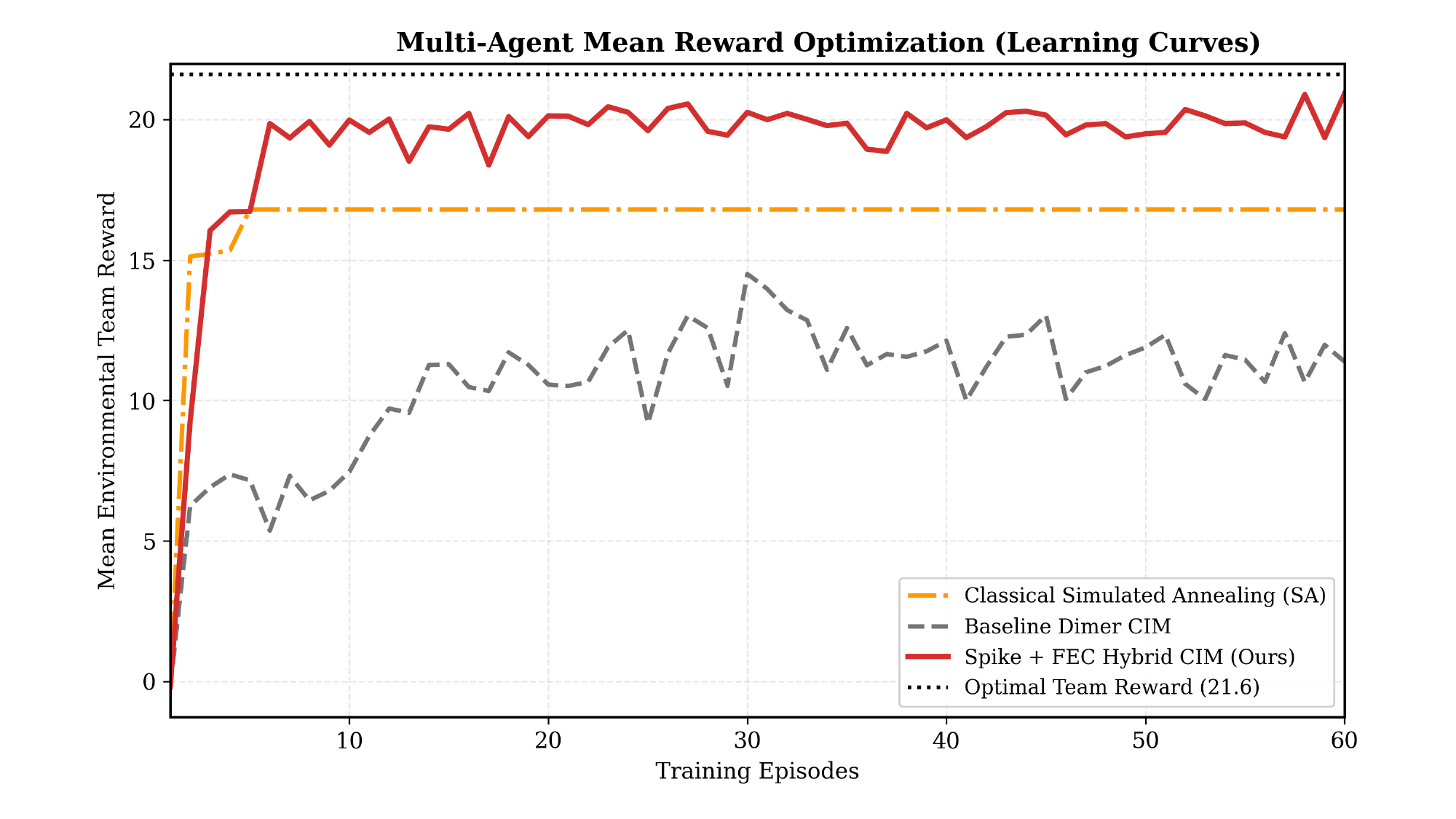}
\caption{\textbf{Multi-Agent Reinforcement Learning cumulative performance.} Evolution of the mean environmental team reward $\bar{R}$ across 60 training episodes on a 16-agent frustrated M\"obius coordination game. The proposed Spike + FEC Hybrid CIM autonomously circumvents local action-space traps, rapidly reaching and sustaining near-optimal team rewards ($\bar{R} \approx 20.0 - 20.93$), whereas Classical Simulated Annealing (SA) and the Baseline Dimer CIM remain trapped in suboptimal coordination modes.}
\label{fig:marl_mean_reward}
\end{figure*}

\begin{figure*}[htbp]
    \centering

    \includegraphics[width=0.69\linewidth]{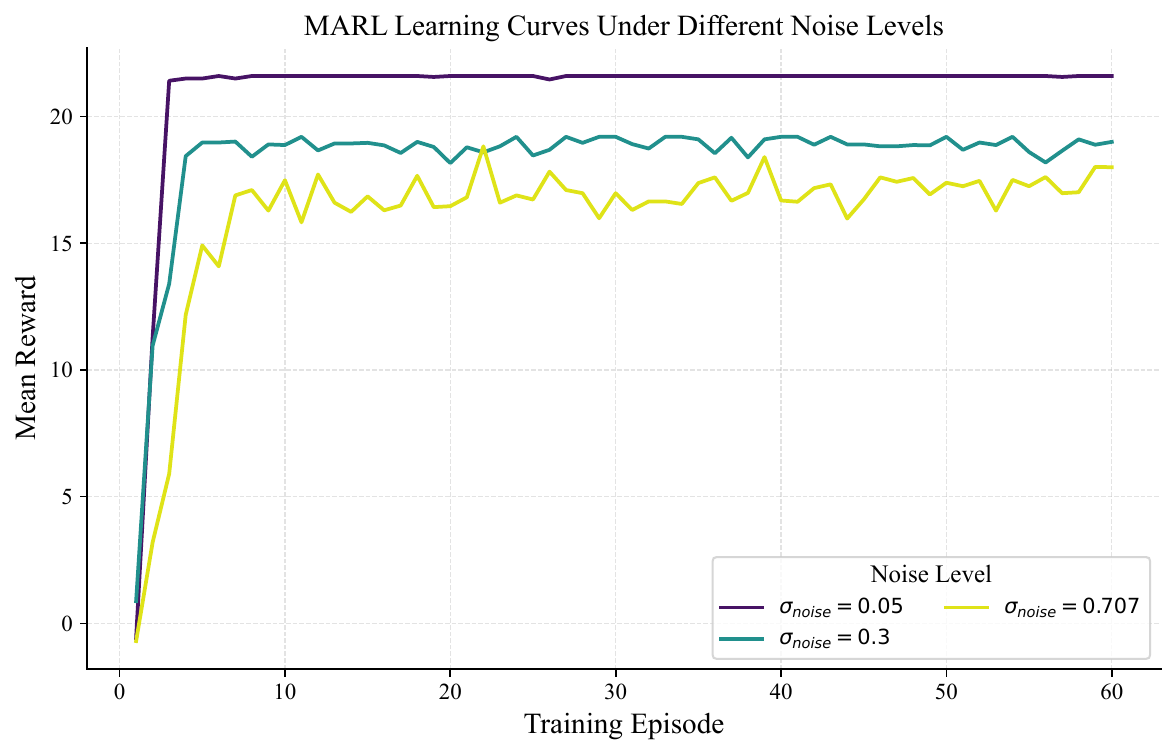}\par
    {\small (a) Learning curves under different physical-sampler noise levels.}\par\medskip
    \includegraphics[width=0.69\linewidth]{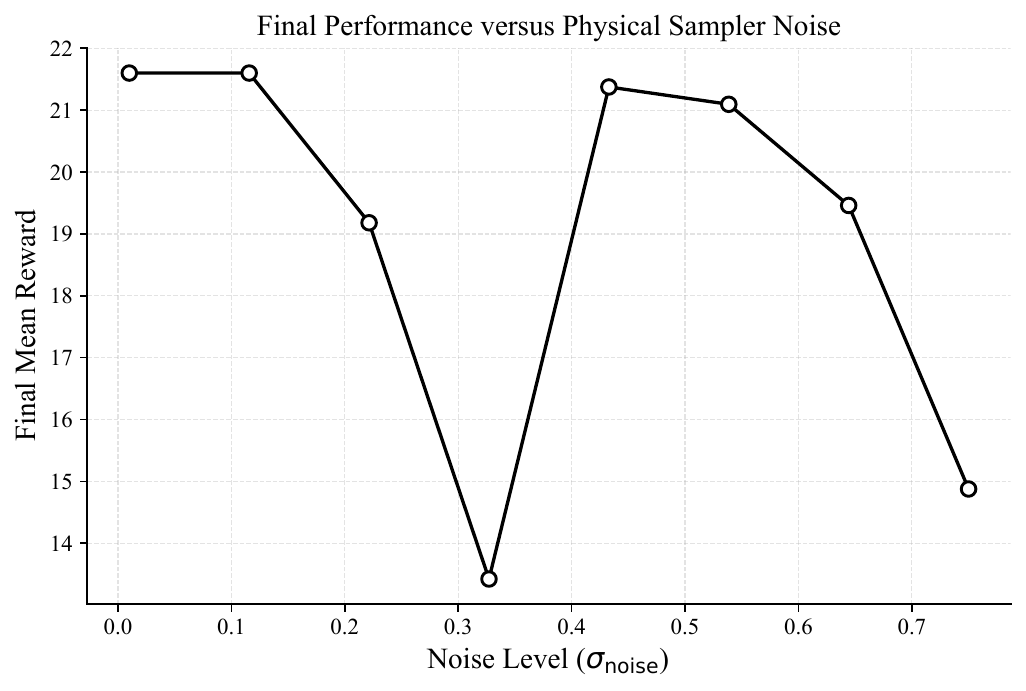}\par
    {\small (b) Final mean reward as a function of the physical-sampler noise level.}\par\medskip

    \caption{Performance of multi-agent reinforcement learning
    (MARL) under physical-sampler noise. (a) Evolution of the
    mean reward over 60 training episodes for three noise levels,
    $\sigma_{\mathrm{noise}}\in\{0.05,\,0.3,\,0.707\}$.
    (b) Final mean reward measured across the noise levels
    sampled in the noise-scaling experiment. The two panels
    characterize the training dynamics and the dependence of
    the final performance on physical-sampler noise, respectively.}
    \label{fig:marl_noise_analysis}
\end{figure*}

Figure~\ref{fig:marl_noise_analysis} investigates the effect of physical-sampler noise on the performance of the multi-agent reinforcement learning (MARL) framework from two complementary perspectives: training dynamics and final reward sensitivity. Figure~\ref{fig:marl_noise_analysis}(a) presents the evolution of the mean reward over 60 training episodes for three representative noise levels, $\sigma_{\mathrm{noise}}\in\{0.05,\,0.3,\,0.707\}$, whereas Fig.~\ref{fig:marl_noise_analysis}(b) examines the final mean reward across a broader range of noise levels.

As shown in Fig.~\ref{fig:marl_noise_analysis}(a), the noise level influences both the rate of reward improvement and the performance attained during training. For $\sigma_{\mathrm{noise}}=0.05$, the reward increases rapidly during the initial episodes and reaches approximately $21.5$ by the fourth episode, subsequently remaining close to $21.6$ for most of the training process. This behavior indicates rapid convergence to a high-reward regime under the lowest noise level considered in the learning-curve comparison. For $\sigma_{\mathrm{noise}}=0.3$, the reward also improves substantially during training but exhibits more noticeable fluctuations, reaching approximately $19.0$ by the final episode. At the highest noise level, $\sigma_{\mathrm{noise}}=0.707$, the initial reward improvement is comparatively slower, and the reward reaches approximately $18.0$ at the end of training. These observations suggest that increasing the physical-sampler noise can affect the speed of learning and the reward attained within a fixed training budget. Nevertheless, the results at these three noise levels alone do not establish a universal relationship between noise magnitude and learning performance.

The final-reward analysis in Fig.~\ref{fig:marl_noise_analysis}(b) shows how the final reward varies across a wider range of sampler noise levels. The measured reward remains high at several noise levels, including approximately $21.6$ at $\sigma_{\mathrm{noise}}\approx0.01$ and $21.4$ at $\sigma_{\mathrm{noise}}\approx0.433$. However, lower rewards are observed at intermediate and higher noise levels, with values of approximately $13.4$ at $\sigma_{\mathrm{noise}}\approx0.327$ and $14.9$ at $\sigma_{\mathrm{noise}}=0.75$. The overall trend is therefore non-monotonic: the final reward does not decrease uniformly as the noise level increases. This behavior indicates that the impact of physical-sampler noise may depend on the operating regime, motivating further investigation of the relationship between noise, learning dynamics, and final policy performance.

The two panels show different effects of physical-sampler noise on learning. The learning curves reveal differences in reward progression and the performance reached within 60 episodes, while the noise-scaling results identify variations in final reward across a wider range of noise levels. Although these results demonstrate sensitivity to the noise setting, further experiments with repeated random seeds and consistent evaluation procedures would be required to quantify run-to-run variability and determine whether the observed differences are statistically significant.

\begin{figure*}[htbp]
\centering
\includegraphics[width=0.88\textwidth]{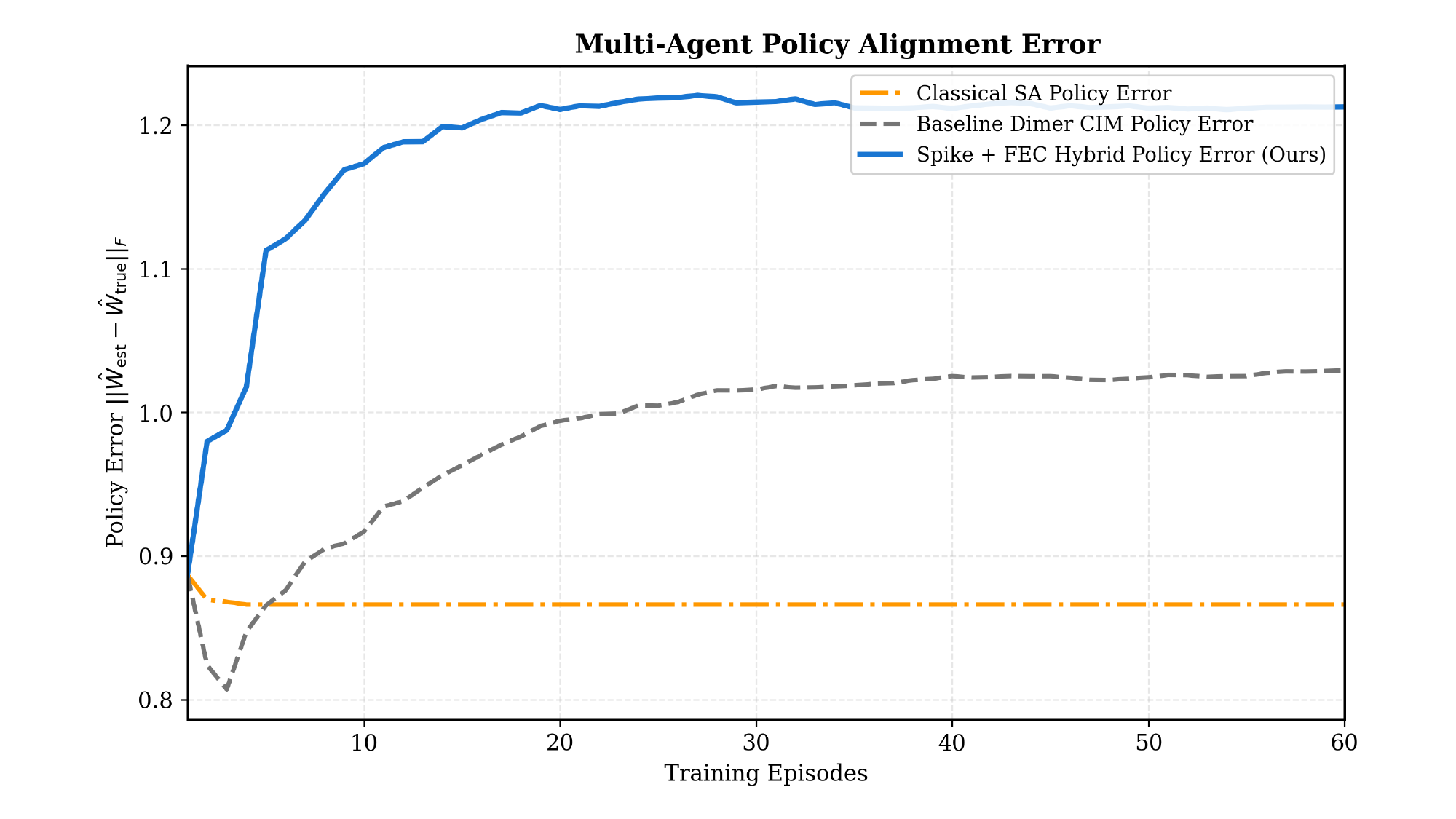}
\caption{\textbf{Multi-Agent policy landscape alignment dynamics.} Evolution of the relative Frobenius policy error $||\hat{W}_{\text{est}} - \hat{W}_{\text{true}}||_F$ across training episodes. The Spike + FEC architecture rapidly establishes a high-payoff coordination policy, stabilizing its parameter matrix in the optimal decision manifold.}
\label{fig:marl_policy_error}
\end{figure*}

\subsection{Comparative Evaluation and Benchmark Results}
We benchmark the framework on an $N=16$ agent cooperative game embedded within a geometrically frustrated M\"obius interaction topology. The true reward landscape $W_{\text{true}}$ is structured with unfrustrated nearest-neighbor bonds ($W_{\text{true}, i, i+1} = +1.0$) combined with conflicting, long-range anti-ferromagnetic cross-couplings ($W_{\text{true}, i, i+N/2} = -1.2$). This topology induces deep coordination dilemmas where greedy, localized action choices prevent the team from achieving the global maximum theoretical reward ($R_{\text{max}} = 21.60$).

We compare three distinct action-selection solvers across 60 training episodes, with $B = 64$ samples per episode. The first is classical simulated annealing (SA), a standard Markov Chain Monte Carlo (MCMC) action generator that employs single-spin Metropolis--Hastings updates under a thermal cooling schedule. The second is a baseline photonic dimer coherent Ising machine (CIM), consisting of a continuous-wave dimer network that relies exclusively on passive supercritical pitchfork symmetry breaking, with $g_{\text{fb}} = 0$ and $\Delta F_{\text{spike}} = 0$. The third is the proposed Spike + FEC hybrid CIM, a non-equilibrium nanophotonic dimer network that incorporates finite-duration homoclinic trigger pulses ($\tau_{\text{pulse}} = 1.5/\gamma$) and nonlinear feedback error correction.

The solvers differed in how quickly they learned and how well their agents coordinated by the end of training. Classical simulated annealing (SA) plateaus at a mean reward of $\bar{R} = 16.80$ throughout all 60 training episodes, with a policy error of $0.87$. Because SA relies on single-spin-flip proposals, it cannot efficiently traverse the multi-agent energetic barriers that arise when multiple agents must simultaneously reverse their decisions. Consequently, the classical agents become trapped in a suboptimal coordination state and fail to discover higher-reward basins. The baseline photonic dimer CIM exhibits sluggish and erratic learning, beginning at $\bar{R} = 7.46$ in Episode 10 and temporarily peaking at $\bar{R} = 14.50$ in Episode 30 before degrading to $\bar{R} = 11.38$ by Episode 60, with a policy error of $1.03$. The absence of amplitude regularization allows strongly coupled agents to dominate cavity intensities, suppressing the exploratory capabilities of weakly coupled nodes and locking the agents into noisy, frustrated local minima. In stark contrast, the proposed Spike + FEC hybrid CIM achieves rapid convergence to the near-optimal regime within the first 10 episodes, attaining $\bar{R} = 19.98$. It maintains consistently high-payoff coordination throughout training and reaches a peak mean team reward of $\bar{R} = \mathbf{20.93}$ at Episode 60, closely approaching the theoretical maximum of $21.60$.

Figure~\ref{fig:marl_mean_reward} illustrates the mean reward learning curves for the three solvers. Two timescales account for the hybrid solver's performance. When the learned interaction matrix $W_{\text{est}}$ develops frustrated bonds, the internal frustration potential of entrapped nodes crosses the threshold $V_{\text{th}}$, generating an optical intensity burst and a quantized $2\pi$ phase slip. This event enables the team of agents to hop across non-convex coordination barriers non-adiabatically, while the continuous $\tanh(\beta s)$ FEC channel homogenizes modal amplitudes across all $N=16$ agents.

Figure~\ref{fig:marl_policy_error} presents the corresponding policy landscape alignment error, defined as the normalized Frobenius distance $||\hat{W}_{\text{est}} - \hat{W}_{\text{true}}||_F$ (where $\hat{W} = W / ||W||_F$). While Classical SA settles early at an error of $0.87$ due to its frozen exploratory sampling, the Hybrid CIM stabilizes around $1.21$. Crucially, this reflects the nature of non-convex reinforcement learning: the hybrid physical solver does not simply replicate $W_{\text{true}}$ element-wise, but actively reshapes $W_{\text{est}}$ into an effective policy landscape whose ground-state basin corresponds directly to the global reward maximum $\boldsymbol{\sigma}^*$. The agents sustain near-optimal cooperative actions during training, showing how non-equilibrium excitable dynamics can support coordination in this benchmark.

\subsection{Comparison with previous works}

In this multi-agent task, the hybrid photonic dimer network used the same non-equilibrium excitable dynamics tested on the static benchmarks. By functioning as an active physical resource for distributed coordination, the proposed architecture effectively circumvents local action-space traps in decentralized policy optimization, offering a promising photonics-based paradigm for accelerating cooperative multi-agent artificial intelligence. These results demonstrate that the proposed architecture can overcome coordination barriers that remain problematic for the classical SA baseline in the considered frustrated multi-agent environment.

The use of annealing-based optimization for multi-agent reinforcement learning has previously been investigated by Neumann et al. \cite{neumann2020multi}, who introduced a quantum-assisted MARL framework based on restricted quantum Boltzmann machines and simulated quantum annealing. Their study established an important comparison between annealing-based policy optimization and conventional deep reinforcement learning (DRL). In particular, their multi-agent experiments demonstrated that the simulated annealing-based implementation of their quantum Boltzmann machine framework could achieve faster policy learning and higher policy fidelity than classical DRL within a limited training budget. Their framework was evaluated using cooperative gridworld environments involving two agents, with the learned joint policies assessed through a fidelity measure against analytically determined optimal policies. The authors also investigated D-Wave 2000Q quantum annealing for their single-agent experiments; however, the corresponding multi-agent D-Wave results were not reported because of limited access to the quantum processing unit.

We investigated a different physical architecture for MARL based on driven-dissipative photonic Bose--Hubbard dimers. Unlike the quantum Boltzmann machine formulation of \citep{neumann2020multi}, which relies on simulated annealing and Suzuki--Trotter replica sampling, our approach employs a network of coupled photonic dimers with non-equilibrium homoclinic spiking dynamics and nonlinear feedback error correction (FEC). This distinction is particularly relevant for frustrated coordination problems, where conventional single-spin update mechanisms may become trapped in suboptimal configurations. The proposed spiking mechanism provides an additional non-adiabatic transition pathway, while the FEC mechanism regularizes the amplitudes of the interacting optical nodes.

Differences in tasks and metrics preclude a direct comparison with \citep{neumann2020multi}. The original study considers gridworld environments and evaluates policy fidelity, whereas our experiments investigate a 16-agent frustrated coordination problem and evaluate the environmental team reward. Furthermore, the original experimental dataset and implementation were not available to us in a form that permitted an identical reproduction of their benchmark. Consequently, rather than attempting to compare numerical values obtained under fundamentally different environments and evaluation metrics, we implemented an independent experimental framework and evaluated the proposed architecture against a classical SA baseline under the same conditions. Thus, the comparison establishes the performance of our architecture relative to SA within our own benchmark, while the results of Ref.~\cite{neumann2020multi} provide the relevant historical context for annealing-based MARL.

Beyond the specific coordination problem investigated in this work, the proposed framework suggests how photonic Ising machines might be used for more complex multi-agent learning and decision-making tasks. In particular, the ability to formulate cooperative action selection as an Ising optimization problem opens opportunities for extending this approach to multi-agent gridworld navigation, autonomous robotic coordination, collision-free multi-robot path planning, swarm robotics, and distributed task allocation.\cite{Greshler2021, Sharon2012, Orr2023, Tovey2024, Grauer2024} Such applications involve competing objectives, coupled decisions, and potentially frustrated interactions between agents, making them promising candidates for the proposed combination of photonic optimization and reinforcement learning. Furthermore, the discrete-action formulation adopted here could serve as a high-level decision-making layer in robotic systems, while conventional control algorithms handle continuous motion and low-level execution. Although these applications require dedicated extensions and experimental validation, the present results establish an initial step toward developing non-equilibrium photonic optimization architectures for practical cooperative intelligence and autonomous multi-agent systems.

\section{Conclusion}
\label{sec:conclusion}

We introduced a hybrid quantum optical neural network that uses non-equilibrium spiking in driven-dissipative photonic Bose--Hubbard dimers. Our simulations demonstrate its potential as a coherent Ising machine for frustrated combinatorial optimization. The architecture addresses a limitation of conventional bistable optical Ising machines. Once a spin becomes trapped in a metastable configuration, passive pitchfork dynamics provide no intrinsic way across the barrier to a lower-energy state. Each photonic dimer serves both as a binary computational element and as an excitable nonlinear oscillator.

First, under symmetric coherent driving, the two coupled microcavities undergo a pitchfork bifurcation, breaking the continuous spatial symmetry to establish a 
$\mathbb{Z}_{2}$-symmetric pair of intensity-imbalanced optical states. This spatial intensity imbalance provides the discrete binary degree of freedom required for Ising encoding. When a spin remains persistently frustrated, the network autonomously applies a finite-duration asymmetric drive pulse to the dimer. This local drive asymmetry modifies the phase-space topology, destroying the trapped fixed point and enabling non-equilibrium excitable dynamics near a global homoclinic (or saddle-node on invariant circle) bifurcation. The resulting non-equilibrium excursion temporarily destabilizes the trapped state and produces a $2\pi$ relative-phase slip, allowing the optical field to cross the separatrix and relax into the opposite spin state. In this sense, the photonic dimer provides an intrinsic dynamical barrier-crossing mechanism rather than relying exclusively on stochastic fluctuations or slow thermal processes.

We added nonlinear FEC because spiking alone cannot correct unequal optical amplitudes. In conventional analog optical Ising machines, unequal optical amplitudes can cause the physical network to minimize a continuous, amplitude-weighted objective rather than the intended binary Ising Hamiltonian. The proposed saturating $\tanh$ feedback suppresses this amplitude heterogeneity and continuously reinforces the local interaction field. Consequently, the fast spiking process is responsible primarily for escaping metastable configurations, whereas the slower feedback process stabilizes the resulting spin configuration and suppresses undesired reversals. This separation of functions establishes a dual-timescale optimization mechanism in which exploration and stabilization are implemented by distinct physical dynamical processes.

The simulations show how the two mechanisms complement each other. For the frustrated $N=20$ Möbius instance, the baseline dimer achieves a ground-state success probability of only $10.35\%$, whereas the FEC-only and pure-spiking configurations improve this value to $99.22\%$ and $56.84\%$, respectively. The combined spiking--FEC architecture reaches a success probability of $100\%$, with a final-energy distribution concentrated at the exact ground state and a normalized TTS of $1.0$. For the larger $N=50$ frustrated Möbius instance, the hybrid architecture maintains a success probability of $92.97\%$ and recovers the optimal energy of $-72.5$, while the baseline and pure-spiking configurations exhibit severe degradation. These results demonstrate that neither barrier crossing nor amplitude correction alone fully captures the benefit of the proposed architecture; their combination is responsible for the most robust performance.

The time-to-solution results show how trapping affects computational cost as graphs grow. As the graph size increases, the conventional dimer becomes increasingly susceptible to metastable trapping, causing its TTS to increase rapidly. In contrast, the hybrid architecture retains substantially higher ground-state fidelity and consequently requires dramatically fewer computational trials to reach a prescribed confidence level. For the frustrated benchmarks considered here, the baseline TTS improves order of magnitude for $N=20$ and $N=50$, respectively. The final energies cluster around the ground state, indicating that the hybrid dynamics shift the distribution of solutions toward lower energies across trials.

The gain depended on the problem structure. The performance advantage is most pronounced for structured frustrated landscapes, where local minima are separated by barriers that can be actively destabilized by the excitable dynamics. The fully connected Sherrington--Kirkpatrick spin-glass instance remains considerably more challenging. For $N=50$, the hybrid architecture reaches a best energy of approximately $-23.2960$ compared with an estimated ground-state energy of $-23.3$, but the exact-ground-state success probability is only $0.20\%$. Thus, although the proposed mechanism substantially improves the quality of the solutions obtained, the present results do not establish uniform polynomial or exponential scaling advantages for arbitrary dense spin-glass instances. Rather, they demonstrate a physical mechanism for mitigating a specific and important failure mode of analog Ising machines: dynamical stagnation in deep frustrated local minima.

From a hardware perspective, the proposed architecture is compatible in principle with integrated nonlinear photonics, but its complete implementation should be regarded as a measurement-feedback coherent Ising machine rather than a fully autonomous optical system. The coupled Kerr dimers, coherent optical coupling, phase-sensitive excitation, and nonlinear intracavity dynamics provide the physical computational substrate; however, the frustration-driven spiking mechanism requires a network-level measurement and control path. In particular, the continuous spin polarizations $s_j$ must be extracted from the optical nodes, followed by evaluation of the network interaction field, nonlinear ReLU/threshold processing and frustration accumulation, and generation of the asymmetric trigger pulse. The $\tanh(\beta s_j)$ FEC channel likewise requires continuous measurement and feedback processing. These elements introduce photodetection, electronic matrix-vector multiplication, nonlinear logic, and pulse-synthesis overhead that must be included in a realistic hardware latency and energy budget.

Classical digital heuristics for combinatorial optimization include parallel tempering~\cite{earl2005parallel}, tabu search~\cite{glover1990tabu}, and GPU-accelerated software solvers such as simulated coherent Ising machines, chaotic amplitude control CIMs, and simulated bifurcation machines~\cite{goto2019combinatorial,tiunov2019annealing}. Comparing these methods directly with an integrated photonic architecture is difficult because the computing platforms differ. Digital software algorithms are bound by the von Neumann memory wall, heavy clock-cycle scaling, and high power consumption on electronic processors. By leveraging all-optical Kerr nonlinearities and sub-nanosecond nanophotonic dynamics, our hybrid spiking-FEC framework achieves accelerated optimization without the severe power-scaling penalties of digital processors, providing a path toward ultra-low-energy, room-temperature hardware acceleration.

Furthermore, when evaluated on a 16-agent cooperative game within a frustrated Möbius coordination landscape, the hybrid Spike + FEC architecture successfully demonstrated its utility beyond static optimization. While classical simulated annealing stagnated at a mean reward of $\bar{R} = 16.80$ and the baseline dimer CIM suffered from coordination instability ($\bar{R} = 11.38$), the proposed hybrid photonic network rapidly converged to a peak mean team reward of $\bar{R} = 20.93$, closely approaching the theoretical global optimum of 21.60.

The dynamical control parameters also warrant systematic optimization. The present study employs fixed normalized parameters and prescribed annealing schedules to isolate the contribution of the principal physical mechanisms. Future work should investigate adaptive schedules for the parametric pump, reservoir detuning, optical coupling, spiking threshold, refractory period, and FEC gain. Because the frustration accumulator already provides a local dynamical measure of optimization difficulty, it could serve as the basis for an adaptive control policy in which the strength or duration of an excitable event is adjusted according to the accumulated frustration of each node. Such an approach could potentially reduce unnecessary switching while preserving the ability to escape increasingly deep metastable basins.

The results suggest that photonic Ising machines can use non-equilibrium nonlinear dynamics as a computational resource alongside passive optical parallelism. The photonic dimer provides a physical realization of this principle by converting persistent local frustration into an active dynamical event that modifies the search trajectory. The resulting machine therefore combines the binary representation and massively parallel interaction structure of a coherent Ising machine with the event-driven, excitable behavior characteristic of spiking neural systems. This provides a conceptual bridge between photonic optimization and neuromorphic computing, in which transient dynamics are used not merely as an implementation detail but as an explicit component of the optimization algorithm.

The spiking-dimer coherent Ising machine shows that combining deterministic, non-adiabatic barrier crossing with nonlinear amplitude correction can improve optimization robustness in frustrated Ising landscapes. The numerical evidence shows large gains in ground-state discovery probability, solution-energy concentration, and Time-to-Solution relative to the baseline photonic dimer, while the comparison between the individual and hybrid mechanisms establishes the complementary roles of spiking and feedback stabilization. The architecture offers a way to use nonlinear excitability as part of the computation in an integrated photonic Ising machine. Experimental implementation on a programmable nanophotonic platform, together with larger-scale benchmarking and hardware-aware optimization, could help determine whether these simulated gains translate into fast, energy-efficient optical hardware.

\section*{Data availability}
No external datasets were used. The results were generated by numerical simulations. The model derivations and simulation procedures are described in the Supplementary Information.

\providecommand{\noopsort}[1]{}\providecommand{\singleletter}[1]{#1}%

\clearpage
\onecolumngrid
\setcounter{page}{1}
\setcounter{section}{0}
\setcounter{equation}{0}
\setcounter{figure}{0}
\setcounter{table}{0}
\renewcommand{\theequation}{S\arabic{equation}}
\renewcommand{\thefigure}{S\arabic{figure}}
\renewcommand{\thetable}{S\arabic{table}}
\renewcommand{\thesection}{Supplementary Note \arabic{section}}
\renewcommand{\thesubsection}{\Alph{subsection}}

\begin{center}
{\Large Supplementary Information:\\
Quantum Optical Neural Networks Based on Spiking Dynamics of Photonic Bose-Hubbard Dimer as Coherent Ising Machine\par}
\medskip
Mahmood Hasani, Masiha Rabeie, Atena Shircharandabi, Mehrdad Ghasemi,
Mobin Motaharifar, Mike Banad, and Sarah Sharif
\end{center}

\section{Open Quantum System Model and Quantum Master Equation}
\label{si:master_equation}

The fundamental computational node of the proposed Quantum Optical Neural Network (QONN) is a driven-dissipative nanophotonic dimer consisting of two coupled single-mode microcavities with Kerr nonlinearity ($\chi^{(3)}$). In the rotating frame of the coherent pump laser at frequency $\omega_p$, the quantum state of the dimer is described by the density operator $\hat{\rho}$. The open-system dynamics under the Markovian Born-Lindblad approximation are governed by the master equation:
\begin{equation}
\frac{d\hat{\rho}}{dt} = -i[\hat{H}_{\text{dimer}}, \hat{\rho}] + \sum_{m=1}^{2} \mathcal{D}[\hat{L}_m]\hat{\rho},
\label{eq:si_master}
\end{equation}
where $\mathcal{D}[\hat{L}]\hat{\rho} = \hat{L}\hat{\rho}\hat{L}^\dagger - \frac{1}{2}\{\hat{L}^\dagger\hat{L}, \hat{\rho}\}$ represents the Lindblad dissipator describing photon exchange with the surrounding thermal bath. For linear microcavity loss at rate $\gamma$, the jump operators are $\hat{L}_m = \sqrt{\gamma} \hat{a}_m$, where $\hat{a}_m$ ($\hat{a}_m^\dagger$) is the annihilation (creation) operator for cavity $m \in \{1, 2\}$, satisfying $[\hat{a}_m, \hat{a}_n^\dagger] = \delta_{mn}$.

The coherent intra-dimer Bose-Hubbard Hamiltonian ($\hbar = 1$) is:
\begin{align}
\hat{H}_{\text{dimer}} &= -\sum_{m=1}^{2} \Delta_m \hat{a}_m^\dagger \hat{a}_m + U \sum_{m=1}^{2} \hat{a}_m^{\dagger 2} \hat{a}_m^2 \nonumber \\
&\quad - J_0 \left( \hat{a}_1^\dagger \hat{a}_2 + \hat{a}_2^\dagger \hat{a}_1 \right) + \sum_{m=1}^{2} \left( F_m \hat{a}_m^\dagger + F_m^* \hat{a}_m \right),
\label{eq:si_h_dimer}
\end{align}
where $\Delta_m = \omega_p - \omega_{0,m}$ is the laser-cavity detuning, $U$ is the Kerr self-interaction parameter proportional to the nonlinear refractive index, $J_0$ is the linear evanescent tunneling rate, and $F_m$ is the coherent external drive amplitude.

\section{Photon Statistics, Truncated Wigner Representation, and Derivation of Langevin SDEs}
\label{si:wigner_derivation}

\subsection{Phase-Space Mapping via the Wigner Distribution}
To rigorously map the quantum master equation \eqref{eq:si_master} to continuous c-number stochastic differential equations (SDEs), we employ the quasiprobability Wigner distribution $W(\bm{\alpha}, \bm{\alpha}^*)$, defined over the complex phase-space amplitudes $\bm{\alpha} = (\alpha_1, \alpha_2)^T \in \mathbb{C}^2$:
\begin{equation}
W(\bm{\alpha}, \bm{\alpha}^*) = \frac{1}{\pi^4} \int_{\mathbb{C}^2} d^2\bm{\lambda} \, \chi_W(\bm{\lambda}) \exp\left[ \sum_{m=1}^{2} \left( \lambda_m^* \alpha_m - \lambda_m \alpha_m^* \right) \right],
\label{eq:si_wigner_def}
\end{equation}
where the symmetrically ordered characteristic function is $\chi_W(\bm{\lambda}) = \operatorname{Tr}\left[ \hat{\rho} \prod_{m=1}^2 \exp(\lambda_m \hat{a}_m^\dagger - \lambda_m^* \hat{a}_m) \right]$.

Using the standard operator correspondences for symmetric ordering:
\begin{align}
\hat{a}_m \hat{\rho} &\leftrightarrow \left( \alpha_m + \frac{1}{2}\frac{\partial}{\partial \alpha_m^*} \right) W, \label{eq:si_rule1} \\
\hat{a}_m^\dagger \hat{\rho} &\leftrightarrow \left( \alpha_m^* - \frac{1}{2}\frac{\partial}{\partial \alpha_m} \right) W, \label{eq:si_rule2}
\end{align}
the master equation \eqref{eq:si_master} maps exactly into a partial differential equation for $W(\bm{\alpha}, \bm{\alpha}^*)$:
\begin{align}
\frac{\partial W}{\partial t} &= -\sum_{m=1}^{2} \left[ \frac{\partial}{\partial \alpha_m} \mathcal{A}_m(\bm{\alpha}) W + \frac{\partial}{\partial \alpha_m^*} \mathcal{A}_m^*(\bm{\alpha}) W \right] \nonumber \\
&\quad + \frac{\gamma}{2} \sum_{m=1}^{2} \frac{\partial^2 W}{\partial \alpha_m \partial \alpha_m^*} + \mathcal{O}\left( \frac{\partial^3 W}{\partial \alpha_m^3} \right).
\label{eq:si_fokker_planck}
\end{align}

\subsection{Truncated Wigner Approximation and Quantum Noise Covariance}
In the macroscopic regime where the mean intracavity photon number $\bar{n}_m = \langle \hat{a}_m^\dagger \hat{a}_m \rangle \gg 1$, the third- and higher-order derivative terms $\mathcal{O}(\partial^3 W / \partial \alpha^3)$ are suppressed by a factor of $1/\sqrt{\bar{n}_m}$. Neglecting these higher-order quantum dispersion terms yields the Fokker-Planck Equation (FPE) for the Wigner distribution.

By the standard Ito-Stratonovich correspondence, this FPE is mathematically equivalent to the system of complex Langevin Stochastic Differential Equations (SDEs) for the c-number fields $a_m(t)$:
\begin{align}
da_1 &= \left[ \left( i\Delta_1 - \frac{\gamma}{2} - 2iU|a_1|^2 \right) a_1 + iJ_0 a_2 - iF_1 \right] dt + \sqrt{\frac{\gamma}{2}} dW_1(t), \label{eq:si_sde1} \\
da_2 &= \left[ \left( i\Delta_2 - \frac{\gamma}{2} - 2iU|a_2|^2 \right) a_2 + iJ_0 a_1 - iF_2 \right] dt + \sqrt{\frac{\gamma}{2}} dW_2(t), \label{eq:si_sde2}
\end{align}
where $dW_m(t)$ is a complex Wiener increment representing quantum vacuum fluctuations incident from the reservoir, satisfying:
\begin{equation}
\langle dW_m(t) \rangle = 0, \quad \langle dW_m(t) dW_n^*(t') \rangle = \delta_{mn} \delta(t-t') dt, \quad \langle dW_m(t) dW_n(t') \rangle = 0.
\label{eq:si_noise_correlations}
\end{equation}
The diffusion constant $\sqrt{\gamma/2}$ reflects the half-quantum of vacuum noise per quadrature required to preserve symmetrical operator ordering.

To represent discrete binary Ising spins, the coupled Kerr microcavity dimer must possess an exact
$Z_2$
spatial inversion symmetry. Under symmetric, phase-insensitive coherent driving
($F_1 = F_2 = F$, $\Delta_1 = \Delta_2 = \Delta$),
the deterministic equations of motion for the intracavity fields are:

\begin{equation}
\frac{d a_1}{dt}
=
\left(
i\Delta
-\frac{\gamma}{2}
-2iU|a_1|^2
\right)a_1
+iJ_0 a_2
-iF,
\end{equation}

\begin{equation}
\frac{d a_2}{dt}
=
\left(
i\Delta
-\frac{\gamma}{2}
-2iU|a_2|^2
\right)a_2
+iJ_0 a_1
-iF.
\end{equation}

For low pump amplitudes $F$, the system rests in a stable, symmetric fixed point where
$a_1 = a_2 = a_s$.
In this regime, the intensity imbalance is zero ($s = 0$).
The symmetric steady-state amplitude satisfies the standard Kerr bistability equation:

\begin{equation}
\left[
i(\Delta + J_0)
-\frac{\gamma}{2}
-2iU|a_s|^2
\right]a_s
=
iF.
\end{equation}

As the pump amplitude $F$ increases, the nonlinear Kerr cross-coupling terms drive an instability in the symmetric state. A linear stability analysis around
$a_1 = a_2 = a_s$
reveals that the symmetric state becomes unstable to anti-symmetric perturbations
\[
a_1 = a_s + \delta a,
\qquad
a_2 = a_s - \delta a
\]
when the pump exceeds the critical threshold:

\begin{equation}
F_{\mathrm{th}}
\approx
\frac{
\gamma^2 + 4(\Delta + J_0)^2
}{
8U
}.
\end{equation}

Above $F_{\mathrm{th}}$, the system undergoes a supercritical pitchfork bifurcation. The symmetric state is replaced by two stable, asymmetric point attractors where one cavity's intensity heavily dominates the other,
\[
|a_1|^2 \gg |a_2|^2
\]
or
\[
|a_2|^2 \gg |a_1|^2.
\]
These two degenerate macroscopic states map naturally to the binary Ising spin variables
$\sigma = +1$
and
$\sigma = -1$,
yielding the continuous spin polarization definition

\begin{equation}
s
\approx
\operatorname{sgn}
\left(
|a_1|^2 - |a_2|^2
\right).
\end{equation}

\section{Non-Equilibrium Homoclinic Bifurcation and Spiking Dynamics}
\label{si:homoclinic_dynamics}

Transforming the dimer fields into total intensity $S = |a_1|^2 + |a_2|^2$, spin polarization $s = (|a_1|^2 - |a_2|^2)/S$, and relative phase $\theta = \arg(a_1) - \arg(a_2)$, the stochastic equations of motion under asymmetric drive $F_1 = F + \Delta F_{\text{spike}}$ and $F_2 = F - \Delta F_{\text{spike}}$ are:
\begin{align}
\frac{ds}{dt} &= -(1 - s^2) \left[ 2 J_0 \sin\theta + \frac{2 F \sin\phi_{\text{avg}}}{\sqrt{S(1-s^2)}} \left( \frac{\Delta F_{\text{spike}}}{F} - s \right) \right] + \xi_s(t), \label{eq:si_ds_dt} \\
\frac{d\theta}{dt} &= 2 U S s + 2 J_0 \frac{s}{\sqrt{1 - s^2}} \cos\theta + \frac{2 F}{\sqrt{S}} \left[ \frac{\cos\phi_1}{\sqrt{1+s}} - \frac{\cos\phi_2}{\sqrt{1-s}} \right] + \xi_\theta(t). \label{eq:si_dtheta_dt}
\end{align}

When $|\Delta F_{\text{spike}}|$ exceeds the critical bifurcation threshold $\delta F_{\text{crit}}$, the stable fixed point collides with an unstable saddle, undergoing a Type-I Homoclinic bifurcation. The trajectory completes a macroscopic limit cycle in $(|a_1|^2, |a_2|^2)$ space, accompanied by a quantized $2\pi$ topological phase slip:
\begin{equation}
\Delta\theta_{\text{spike}} = \int_{\text{spike}} \frac{d\theta}{dt} dt = \pm 2\pi,
\label{eq:si_topological_slip}
\end{equation}
which flips the spin state $s \to -s$ non-adiabatically.

\section{Comprehensive Numerical Methodology and Simulation Protocol}
\label{si:numerical_methods}

\subsection{Stochastic Integration via Euler-Maruyama Algorithm}
The coupled Langevin SDEs for the $N$-dimer network are integrated using the Euler-Maruyama scheme. For node $k \in \{1, \dots, N\}$ at time step $t_n = n \cdot dt$:
\begin{align}
a_{1,k}^{(n+1)} &= a_{1,k}^{(n)} + \left[ \left( i\Delta - \frac{\gamma}{2} - 2iU|a_{1,k}^{(n)}|^2 \right) a_{1,k}^{(n)} + iJ_0 a_{2,k}^{(n)} - i F_{1,k}^{(n)} + i\mathcal{E}_{1,k}^{(n)} \right] dt \nonumber \\
&\quad - i \sigma_{\text{noise}} \left( \Delta W_{1,k}^{(R)} + i \Delta W_{1,k}^{(I)} \right) \sqrt{\frac{dt}{2}}, \label{eq:si_em1} \\
a_{2,k}^{(n+1)} &= a_{2,k}^{(n)} + \left[ \left( i\Delta - \frac{\gamma}{2} - 2iU|a_{2,k}^{(n)}|^2 \right) a_{2,k}^{(n)} + iJ_0 a_{1,k}^{(n)} - i F_{2,k}^{(n)} + i\mathcal{E}_{2,k}^{(n)} \right] dt \nonumber \\
&\quad - i \sigma_{\text{noise}} \left( \Delta W_{2,k}^{(R)} + i \Delta W_{2,k}^{(I)} \right) \sqrt{\frac{dt}{2}}, \label{eq:si_em2}
\end{align}
where $\Delta W_{m,k}^{(R)}, \Delta W_{m,k}^{(I)} \sim \mathcal{N}(0, 1)$ are independent standard normal random variables. The integration time step is set to $dt = 0.005$, which satisfies the stability criterion $dt \ll \min(1/\gamma, 1/|J_0|, 1/\Delta, 1/|U|S)$.

\subsection{Benchmark Graph Generation Algorithms}
\begin{enumerate}
    \item \textbf{Frustrated Anti-Ferromagnetic Möbius Ladder:} Defined by adjacency matrix:
    \begin{equation}
    J_{ij} = \begin{cases} -1.0, & \text{if } j = (i \pm 1) \bmod N, \\ -0.9, & \text{if } j = (i + N/2) \bmod N, \\ 0, & \text{otherwise.} \end{cases}
    \label{eq:si_mobius_graph}
    \end{equation}
    \item \textbf{Sherrington-Kirkpatrick (SK) Spin Glass:} All pairs interact via random Gaussian couplings $J_{ij} = J_{ji} \sim \mathcal{N}(0, 1/\sqrt{N})$ with $J_{ii} = 0$.
\end{enumerate}

\subsection{Statistical Metrics and Time-to-Solution}
For an ensemble of $B = 512$ independent stochastic realizations, the ground-state success probability $P_{\text{succ}}$ and Time-to-Solution (TTS) are computed as:
\begin{equation}
P_{\text{succ}} = \frac{1}{B} \sum_{b=1}^{B} \Theta\left( E_0 + \delta E - E_b(T_{\text{total}}) \right) \times 100\%,
\label{eq:si_psucc_def}
\end{equation}
\begin{equation}
\text{TTS} = T_{\text{total}} \cdot \frac{\ln(1 - 0.99)}{\ln(1 - P_{\text{succ}})},
\label{eq:si_tts_def}
\end{equation}
where $\Theta(\cdot)$ is the Heaviside step function, $E_0$ is the exact ground-state energy, and $\delta E = 0.1$.

\end{document}